\documentclass[pdflatex,sn-aps]{sn-jnl}

\usepackage{algorithm}
\usepackage{algorithmicx}
\usepackage{algpseudocode}
\usepackage{amsmath,amssymb,amsfonts}
\usepackage{amsthm}
\usepackage[title]{appendix}
\usepackage{booktabs}
\usepackage{bm}
\usepackage{braket}
\usepackage{cleveref}
\usepackage{graphicx}
\usepackage{listings}
\usepackage{manyfoot}
\usepackage{multirow}
\usepackage{textcomp}
\usepackage{xcolor}

\newcommand{\Ai}{\mathop{\mathrm{Ai}}}
\newcommand{\Bi}{\mathop{\mathrm{Bi}}}

\newcommand{\arcoth}{\mathop{\mathrm{arcoth}}}

\begin{document}

\title[Article Title]{Conditions for Quantum Violation of Macrorealism in Large-spin Limit}

\author[1]{\fnm{Qi-Hong} \sur{Cai}}
\equalcont{These authors contributed equally to this work.}

\author[1]{\fnm{Xue-Hao} \sur{Yu}}
\equalcont{These authors contributed equally to this work.}

\author[1]{\fnm{Ma-Cheng} \sur{Yang}}

\author[1]{\fnm{Ao-Xiang} \sur{Liu}}

\author*[1,2]{\fnm{Cong-Feng} \sur{Qiao}}\email{qiaocf@ucas.ac.cn}

\affil[1]{\orgdiv{School of Physical Sciences}, \orgname{University of Chinese Academy of Sciences}, \orgaddress{\street{YuQuan Road 19A}, \city{Beijing}, \postcode{100049}, \country{China}}}
\affil[2]{\orgdiv{Key Laboratory of Vacuum Physics}, \orgname{CAS}, \orgaddress{\street{YuQuan Road 19A}, \city{Beijing}, \postcode{100049}, \country{China}}}

\abstract{This study investigates the emergence of macroscopic classical behavior from quantum foundations via the entropic Leggett--Garg inequality. We introduce a geometric framework for deriving entropic Leggett--Garg inequalities with higher-order temporal correlations and demonstrate their advantages over conventional formulations. Numerical analyses show that entropic Leggett--Garg inequalities offer a robust and complementary criterion to standard approaches, providing a transparent information theoretic interpretation that facilitates the characterization of coherent quantum processes. By applying the WKB approximation, we prove that violations for maximally mixed states remain bounded by a constant in the macroscopic limit, indicating that macrorealism dominates in generic parameter regimes. We further explain previously reported maximal violations at specific parameter regimes as a consequence of the breakdown of the WKB approximation. Our findings indicate that quantum and classical descriptions remain macroscopically incompatible, while violations persist only in fine-tuned regimes, clarifying the conditions for detecting macroscopic quantum phenomena.}


\maketitle

\section{Introduction}

Quantum mechanics (QM) exhibits counterintuitive features, such as nonlocality \cite{HHHH09} and contextuality \cite{BCG+22}, which are fundamentally incompatible with the classical description of macroscopic systems. Despite the success of QM at atomic and subatomic scales, characterizing the crossover between the quantum and classical regimes remains a fundamental challenge. A central question \cite{FSD+18} is whether quantum phenomena intrinsically vanish with increasing system size, or persist at the macroscopic scale under specific conditions. Although superconductivity \cite{MGC63} and superfluidity \cite{Leg99} are frequently cited as paradigms of ``macroscopic quantum phenomena'', Leggett \cite{Leg80} argued that these are merely macroscopic accumulations of microscopic quantum effects, and hence insufficient to determine whether the quantum mechanical description can be extrapolated to macroscopic systems. Instead, Leggett proposed that the superposition of macroscopically distinguishable states could serve as more convincing experimental evidence.

To this end, a method capable of distinguishing genuine quantum superposition from classical statistical mixtures is needed. Leggett and Garg \cite{LG85,Leg08} introduced a definition of ``classical worldview'' known as macrorealism, consisting of two principal assumptions: (1) Macroscopic realism: A macroscopic system with two or more macroscopically distinct states available to it will at all times be in one or the other of these states; (2) Noninvasive measurability at the macroscopic level: It is possible, in principle, to determine the state of the system with arbitrarily small perturbation on its subsequent dynamics. Analogous to Bell inequalities \cite{Fin82}, Leggett and Garg \cite{LG85} showed that the temporal correlations must obey a class of inequalities under macrorealism, for instance,
\begin{eqnarray}
1+C_{12}+C_{23}+C_{13}\ge0\ ,&\\
|C_{12}+C_{23}+C_{14}-C_{24}|\le2\ ,&
\end{eqnarray}
where $C_{ij}\equiv\braket{Q(t_{i})Q(t_{j})}$ denote the temporal correlation functions. Violations of such Leggett--Garg inequalities (LGIs) sufficiently (not necessarily) indicate the failure of macrorealism.

Such violations have been observed in a variety of experiments employing diverse measurement protocols, ranging from pre-diagonalized measurements \cite{SOS11,ARM11} and weak measurements \cite{PMN+10,DBHJ11,GAB+11} to the more rigorous ideal negative measurements (INM) \cite{Leg88}. The INM approach enforces non-invasiveness via null-result postselection. If no detector click is observed, one infers that the system occupied a state not coupled to the probe, implying that the measurement interaction left the system effectively undisturbed. Implementations of this principle generally fall into two categories: (i) ancilla-assisted schemes \cite{XLZG11,KSG+12,KSRM13}, where the system is entangled with auxiliary systems at the intermediate time and the projective readout is performed only at the end; and (ii) interferometric null-result schemes \cite{ELN12,KWG+24}, which exploit interaction-free measurement principles similar to the Elitzur--Vaidman bomb test \cite{EV93}. Notably, these non-invasive protocols have been successfully extended to multi-outcome scenarios \cite{KBLL17}, demonstrating their feasibility for the high-dimensional systems.

Beyond the conventional application in testing macrorealism, LGIs also serve as indicators of quantum coherence \cite{LECN10,ELN13,ZHLG15,SHLY18,ERT22}, certifying dynamical non-classicality even in scenarios where static, state-based measures \cite{BCP14} prove inadequate. For instance, a maximally mixed state can yield LGI violations during unitary evolution, even though it remains incoherent throughout the process \cite{KB07,MDH16}. To probe such quantum violations across broader physical scenarios, various LGI formulations have been developed \cite{KP17a,KP20}, most notably the standard correlator-based forms \cite{Hal16,Hal19,Hal19a,HM20} and the probability-based Wigner forms \cite{SMPH15,KP17}. Nevertheless, these approaches are typically tailored to  a fixed number of outcomes, thus hindering adaptation to high-dimensional systems or experimental setups with detection inefficiencies \cite{CF12}.

To address these limitations, an information-theoretic approach based on Shannon entropy, known as the entropic Leggett--Garg inequality (ELGI), was introduced \cite{DKSR13,Ras14}. Whereas, hitherto the existing ELGIs are all confined to two-point joint entropies. It is thus of considerable interest to explore ELGIs involving higher-order temporal correlations, which would offer enhanced detection sensitivity and manifest novel phenomena. Furthermore, the precise relationship between this entropic formulation and other established methods, such as standard LGIs (SLGIs) \cite{Hal16,Hal19,Hal19a,HM20} and Wigner form of LGIs (WLGIs)  \cite{SMPH15,KP17}, remains unclear.

In this study, we investigate the persistence of violations of entropic Leggett-Garg inequalities (ELGIs) at the macroscopic scale and their utility in characterizing quantum dynamics. To this end, we introduce a systematic approach to derive ELGIs with higher-order temporal correlations. We then demonstrate the operational advantages of ELGIs, highlighting their complementarity to conventional LGIs and their ability to detect non-Markovian memory effects directly through observable statistics. Furthermore, we analyze ELGI violations in spin-$j$ systems driven by a constant external field and examine their asymptotic behavior in the limit $j \to \infty$ through the WKB approximation. Our analysis reveals that, in the regime where the WKB approximation is valid, quantum violations persist but are negligible relative to the system size; conversely, when the WKB approximation breaks down, the system rapidly reaches the maximal quantum violation of $\ln(2j+1)$. These findings delineate clear boundaries for the experimental testing of entropic Leggett--Garg inequalities at the macroscopic scale.

\section{Entropic Leggett--Garg inequality}\label{sec1}

\begin{figure}[ht]
\centering
\includegraphics[width=\textwidth]{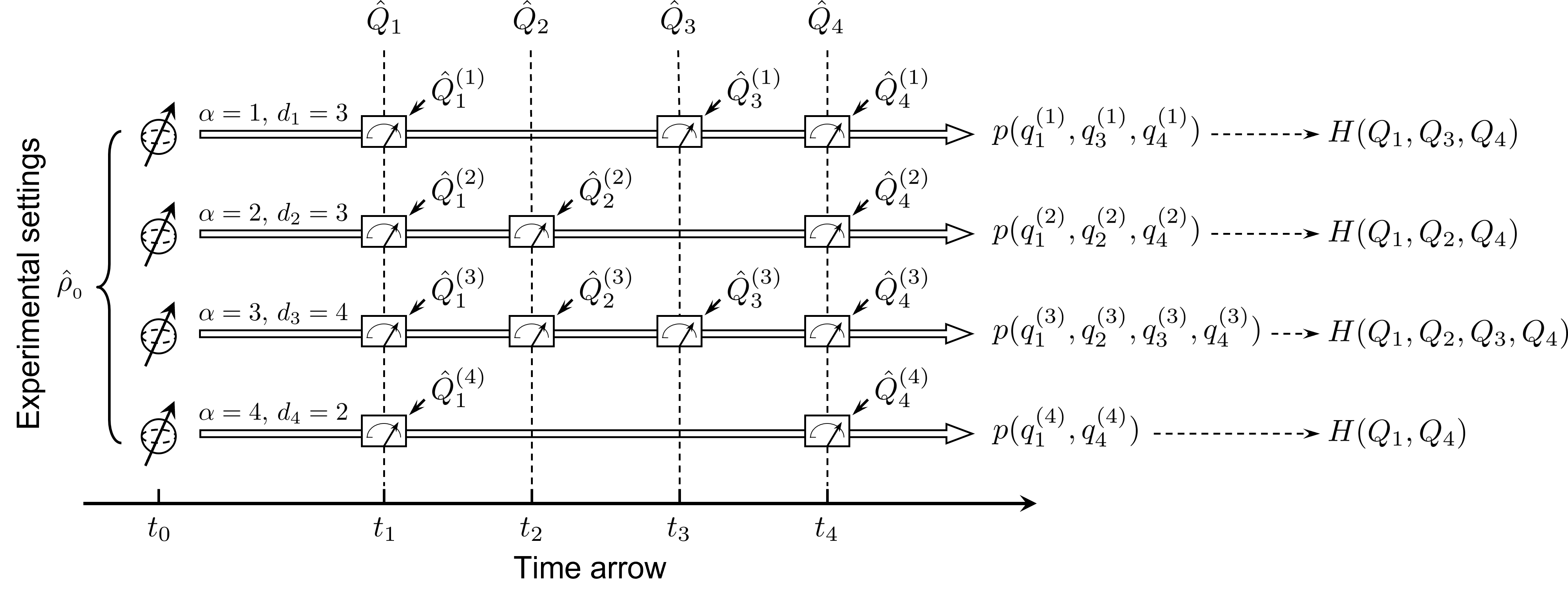}
\caption{{\bfseries Schematic representation of the experimental setting for ELGI.} The elementary type ELGI $\mathcal{D}_{2,3}=H(Q_{1},Q_{3},Q_{4})+H(Q_{1},Q_{2},Q_{4})-H(Q_{1},Q_{2},Q_{3},Q_{4})-H(Q_{1},Q_{4})\ge0$ is taken as an example, where $Q_{i}$ denotes $Q(t_{i})$ for simplicity. All the $M$ experiments begin with the same initial state $\hat{\rho}_{_{0}}$. The joint entropy must be obtained from independent experiments rather than derived from the marginal probability distribution of a single experiment with measurements at all time points, as later measurements could be affected by earlier ones.}\label{fig:schematic}
\end{figure}

We begin by recounting the experimental scheme proposed by Leggett and Garg \cite{LG85} in deriving the LGIs. Given an observable $Q$, which can be treated as a variable mathematically, to measure it in an experiment non-invasively, one performs a series of measurements at some or all points of $n$ potential times $(t_{1},\cdots,t_{n})$. In the $\alpha$-th individual experiment, measurements are conducted at $d$ points of those $n$ times in sequence to acquire multipoint joint probabilities $p\big(q^{(\alpha)}(t_{i}),q^{(\alpha)}(t_{j}),\cdots,q^{(\alpha)}(t_{l})\big)$, while the system remains undisturbed at the other times. Here, $q^{(\alpha)}(t_i)$ denotes the measurement outcome of $Q$ at time $t_i$, the indices $i,j,\cdots,l\in\{1,2,\cdots,n\}$, and $\mathop{\mathrm{card}}\{i,j,\cdots,l\}=d$, as illustrated in \cref{fig:schematic}.

Suppose $M$ rounds of experiments are performed in total, each of which provides a joint entropy
\begin{equation} \label{eqn:Shannon}
H\big(Q^{(\alpha)}(t_{i}),\cdots ,Q^{(\alpha)}(t_{l})\big)=-\sum_{p}{p\big(q^{(\alpha)}(t_{i}),\cdots , q^{(\alpha)} (t_{l})\big)\ln{p\big(q^{(\alpha)}(t_{i}),\cdots , q^{(\alpha)}(t_{l})\big)}}\ .\ 
\end{equation}
Here, $Q^{(\alpha)}(t_{i})$ represents the variable $Q$ being measured in the $\alpha$-th experiment at time $t_i$. The $M$ joint entropies can then be arranged into a vector $\vec{h}_{\text{obs}}\in\mathbb{R}^{M}$. It should be noted that $Q^{(\alpha)}(t_{i})$ and $Q^{(\beta)}(t_{i})$ are distinct random variables due to the potential disturbance induced in measurement, even if they are measured at the same time $t_{i}$. Consequently, there are  $d_{1}+\cdots+d_{M}$ random variables in total. In contrast, macrorealism posits that measurements are non-invasive, such that any measurement merely reveals the pre-existing value $q_{\text{pre}}{(t)}$ at time $t$ without perturbation, which leads to
\begin{equation}\label{eqn:macro-condition}
H\big(Q^{(\alpha)}(t_{i}),\cdots,Q^{(\alpha)}(t_{l})\big)=H\big(Q_{\text{pre}}(t_{i}),\cdots,Q_{\text{pre}}(t_{l})\big)\ ,
\end{equation}
for all $M$ joint entropies. It should be noted that $Q_{\text{pre}}(t_{i})$ remains identical across all $M$ rounds of experiments; consequently, a macrorealistic model involves no more than $n$ random variables. Yeung \cite{Yeu97} proved that entropic vectors $\vec{h}_{\text{obs}}$ satisfying \cref{eqn:macro-condition} are confined within a convex cone $\Gamma_{\text{E}}\subset\mathbb{R}^{M}$. Therefore, observation of $\vec{h}_{\text{obs}}$ lying outside $\Gamma_{\text{E}}$ indicates violation of macrorealism. 

Since a complete characterization of $\Gamma_{\text{E}}$ remains an open problem in information theory, we employ the Shannon cone $\Gamma_{\text{Sh}}$ defined by the non-negativity of conditional entropy and mutual information as an outer approximation, from which we derive two types of entropic Leggett--Garg inequalities:
\begin{align}
\mathcal{D}_{i}:\!&=H(Q_{i}|Q_{\text{All}}-\{Q_{i}\})=H\big(Q_{\text{All}}\big)-H\big(Q_{\text{All}}-\{Q_{i}\}\big)\ge0\ ,\label{eqn:conditional-entropy}\\ 
\mathcal{D}_{i,j}:\!&=I(Q_{i};Q_{j}|Q_{\text{All}}-\{Q_{i},Q_{j}\})\notag\\
&=H\big(Q_{\text{All}}-\{Q_{i}\}\big)+H\big(Q_{\text{All}}-\{Q_{j}\}\big)- H\big(Q_{\text{All}}\big)-H\big(Q_{\text{All}}- \{Q_{i},Q_{j}\}\big)\ge0\ .\label{eqn:conditional-mutual-information}
\end{align}
Here, $Q_{\text{All}}=\{Q_{1},\cdots,Q_{n}\}$ with $Q_{i}$ denoting $Q^{\alpha}(t_{i})$ for simplicity. Yeung \cite{Yeu97} demonstrated that these two types of inequalities provide a complete characterization of the Shannon cone; thus, $\mathcal{D}_{i}$ and $\mathcal{D}_{i,j}$ constitute two fundamental basis elements from which all other Shannon-type ELGIs can be systematically derived. For instance, the $n=3$ case of ELGI proposed by Devi {\itshape et al.} \cite{DKSR13}, though originally derived through a different method, can be expressed as $H(Q_{2}|Q_{1})+H(Q_{3}|Q_{2})-H(Q_{3}|Q_{1})=\mathcal{D}_{1,3}+\mathcal{D}_{2}\ge0$. This geometric approach parallels the method introduced by Chaves and Fritz for entropic Bell and contextuality inequalities \cite{FC13,CF12}.

\begin{figure}[ht]
\centering
\includegraphics[width=0.49\textwidth]{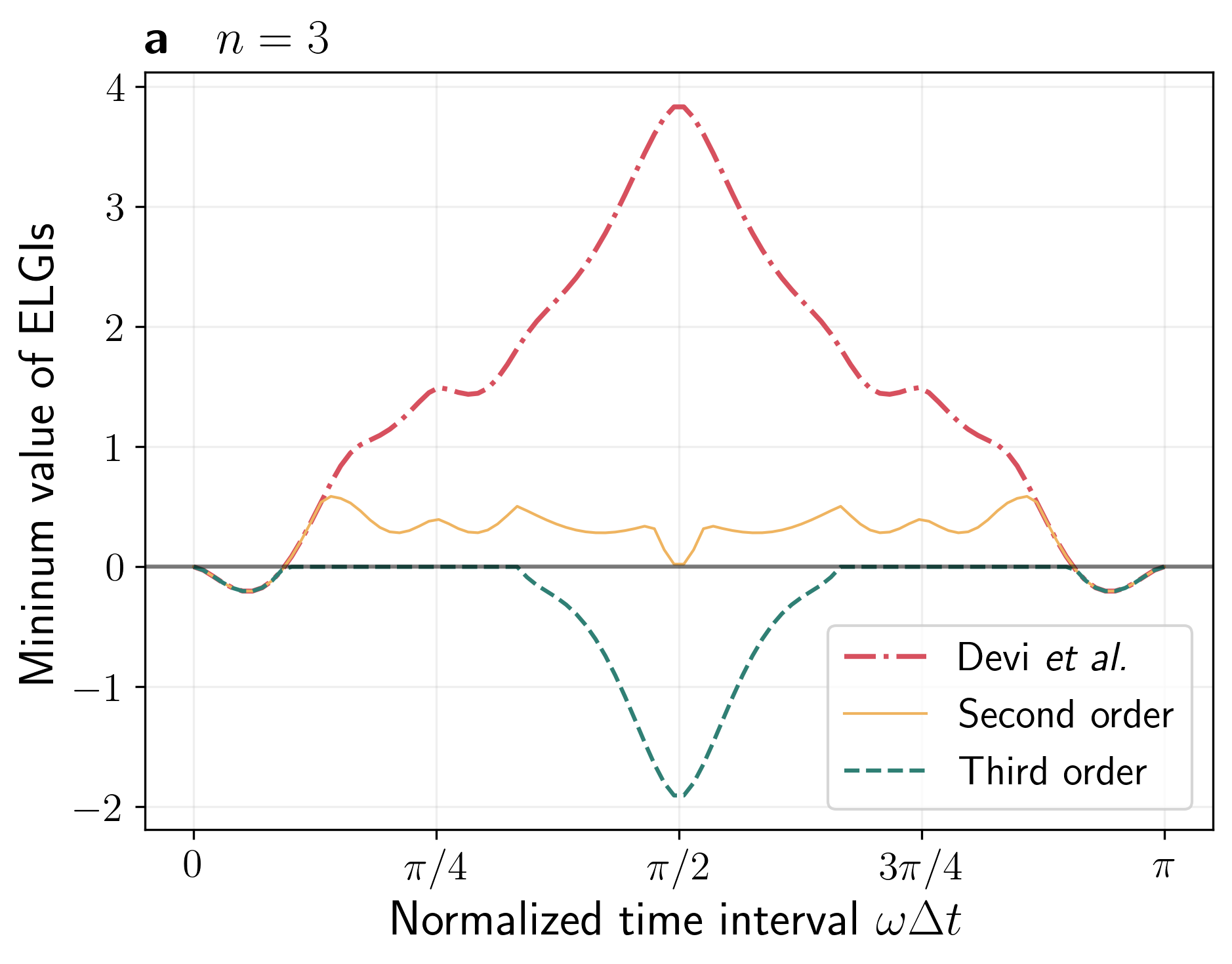}
\includegraphics[width=0.49\textwidth]{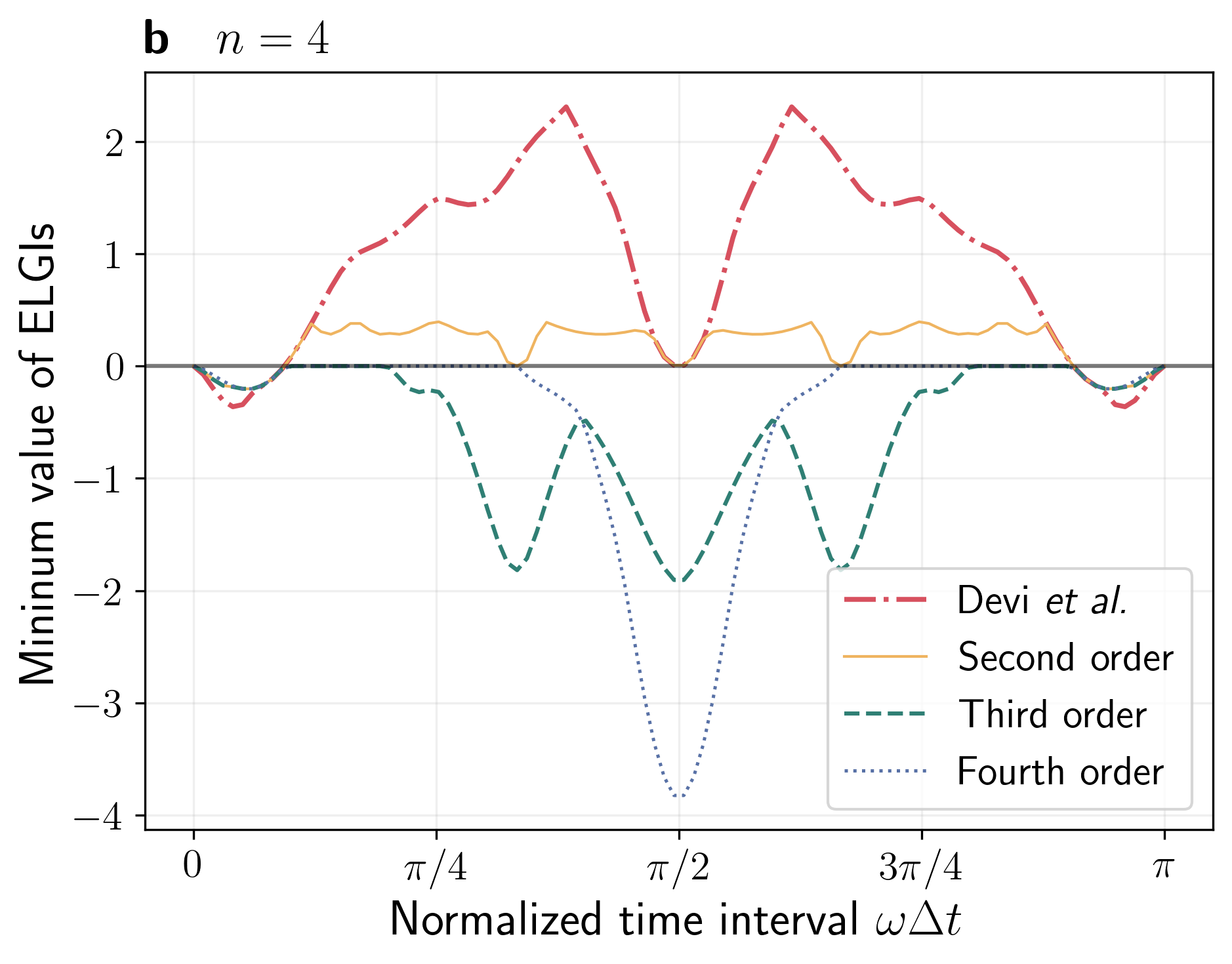}
\caption{{\bfseries Comparison of quantum violations of ELGIs involving different orders of temporal correlations.} All possible Shannon-type ELGIs are grouped according to their order. ELGIs that can be written as sums of other ELGIs of the same order are excluded, since they do not provide any additional information about quantum violations. Numerical calculations are performed for a spin-2 system with Hamiltonian $\hat{H}=\omega\hat{J}_{y}$ and observable $\hat{Q}=\hat{J}_{z}$, considering (a) $n=3$ and (b) $n=4$ equally spaced measurement time points, $t_{i+1}-t_{i}=\Delta t$. The complete list of inequalities involved in the figure is provided in the supplementary material.}\label{fig:comp-order}
\end{figure}

A key advantage of our approach is its capacity to deriving ELGIs in higher order temporal correlation. As shown in \cref{fig:comp-order}, higher-order ELGIs show larger violations and over a broader parameter range, comparing to Devi's second-order ELGIs. Similar enhancements obtained from higher-order correlations have also been observed in other Leggett--Garg criteria, such as standard LGIs \cite{PQK18,Hal19a} and no-signaling in time (NSIT) conditions \cite{MHL21,KP22}. The enhanced sensitivity facilitates more effective detection of quantum behavior in temporal correlations, particularly in systems with intricate dynamics where two-point correlations might fail to capture subtle characteristics, such as non-Markovian processes \cite{TPM+19}. Notably, the measurement of these higher-order correlations can be achieved by extending existing ideal negative-measurement schemes \cite{XLZG11,KSG+12,KSRM13,ELN12,KWG+24}, as illustrated by a simple example in the Supplementary Material.

\section{Advantages of Entropic Leggett--Garg Inequalities}

While ELGIs are derived from elementary information-theoretic constraints, SLGIs and WLGIs originate from linear positivity conditions imposed on an underlying quasiprobability distribution \cite{Hal16,SMPH15}. Consequently, ELGIs cannot be obtained from SLGIs or WLGIs, nor can SLGIs or WLGIs be inferred from ELGIs. This formal distinction is further illustrated through numerical analyses of spin systems.

\begin{figure}[ht]
\centering
\includegraphics[width=0.49\textwidth]{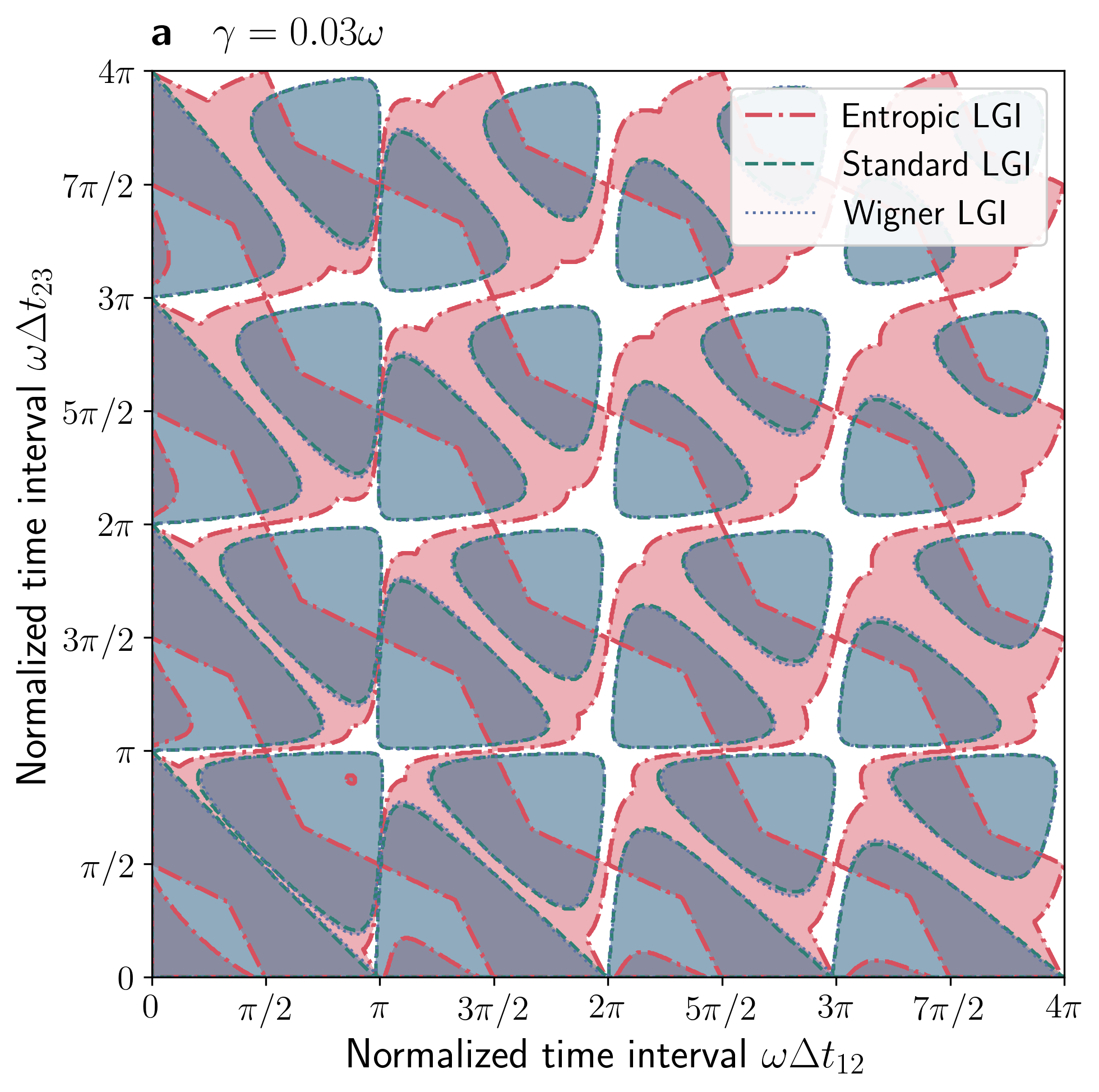}
\includegraphics[width=0.49\textwidth]{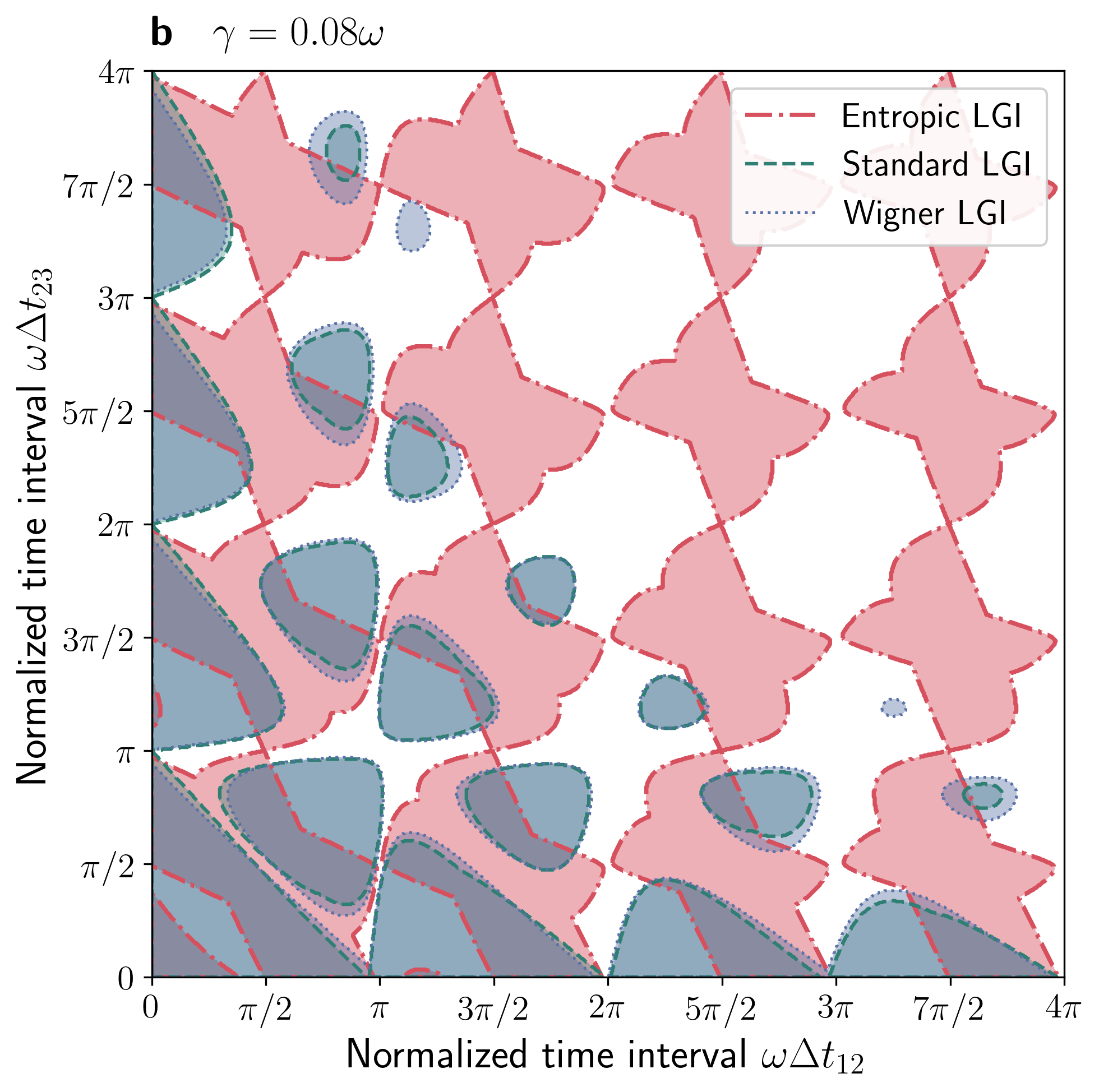}
\caption{{\bfseries Comparison of quantum-violation regions for ELGIs, SLGIs, and WLGIs.} 
Shaded regions indicate parameter regimes where at least one inequality in each family is violated: red for ELGIs, green for SLGIs, and blue for WLGIs. 
Numerical results are obtained for a spin-1 system governed by the Lindblad master equation 
$\mathrm{d}\hat{\rho}/\mathrm{d}t=-i[\omega \hat{J}_{y},\hat{\rho}]+\gamma\big(\hat{J}_{-}\hat{\rho}\hat{J}_{+}-\tfrac{1}{2}\{\hat{J}_{+}\hat{J}_{-},\hat{\rho}\}\big)$ 
with the observable $\hat{Q}=\hat{J}_{z}$ and $n=3$ measurement times. 
The panels correspond to different decay rates: (a) $\gamma=0.03\omega$ and (b) $\gamma=0.08\omega$, demonstrating that ELGI violations persist under stronger dissipation, whereas violations in the other families may disappear.}
\label{fig:comp-lgi}
\end{figure}

As shown in \cref{fig:comp-lgi}, the violation regions of ELGIs, SLGIs, and WLGIs do not exhibit a strict inclusion relation. Although SLGIs and WLGIs display substantially overlapping violation regions owing to their shared algebraic structure, the ELGI violation region overlaps only partially with either family and is largely complementary to both. This complementarity indicates that ELGIs can detect temporal quantum correlations in parameter regimes where conventional LGI tests show no violations. Notably, ELGIs require neither additional experimental observables nor modified measurement protocols, since all three LGI families are evaluated from the same multi-time probabilities obtained using an identical measurement sequence. Thus, ELGIs provide an alternative, information-theoretic consistency test for macrorealism, enhancing detection capability without adding experimental complexity.

Under dissipative evolution, ELGIs maintain detectable violations over longer time intervals and under stronger dissipation compared with SLGIs and WLGIs. This robustness to decoherence makes the entropic test particularly suitable for probing quantum behavior in noisy or open-system scenarios. Further investigation is needed to clarify the mechanisms underlying this robustness.

Furthermore, the ELGI framework offers a clear information-theoretic interpretation, being directly grounded in the non-negativity of conditional entropy and conditional mutual information. This property allows any Shannon-type information equality or inequality to be mapped onto a corresponding ELGI, providing a systematic method to distinguish information disturbances arising from invasive quantum measurements from those due to other physical effects.

As an illustrative application, we employ ELGIs to characterize non-Markovian behavior in quantum dynamics. Specifically, we consider repeated measurements of an observable $\hat{Q}$ at three time points $t_{1}<t_{2}<t_{3}$, yielding outcomes $Q_{1}, Q_{2}, Q_{3}$. In a classical Markov process, the conditional independence implied by Markovianity requires
\begin{equation}\label{eqn:Markovian}
H(Q_{3}|Q_{2})=H(Q_{3}|Q_{1},Q_{2}),
\end{equation}
which corresponds to the ELGI
\begin{equation}
\mathcal{D}_{1,3}=H(Q_{2},Q_{3})+H(Q_{1},Q_{2})-H(Q_{1},Q_{2},Q_{3})-H(Q_{2})\ge0.
\end{equation}
Any deviation from the classical Markovianity condition \cref{eqn:Markovian} has a clear physical interpretation. Positive violations indicate non-Markovian memory effects in the system or environment, whereas negative violations typically arise from the invasiveness of quantum measurements. Therefore, in scenarios where a Markovian approximation predicts no quantum violation, any observed positive deviation serves as a clear signature of genuine non-Markovian dynamics.

\begin{figure}[ht]
\centering
\includegraphics[width=0.42\textwidth]{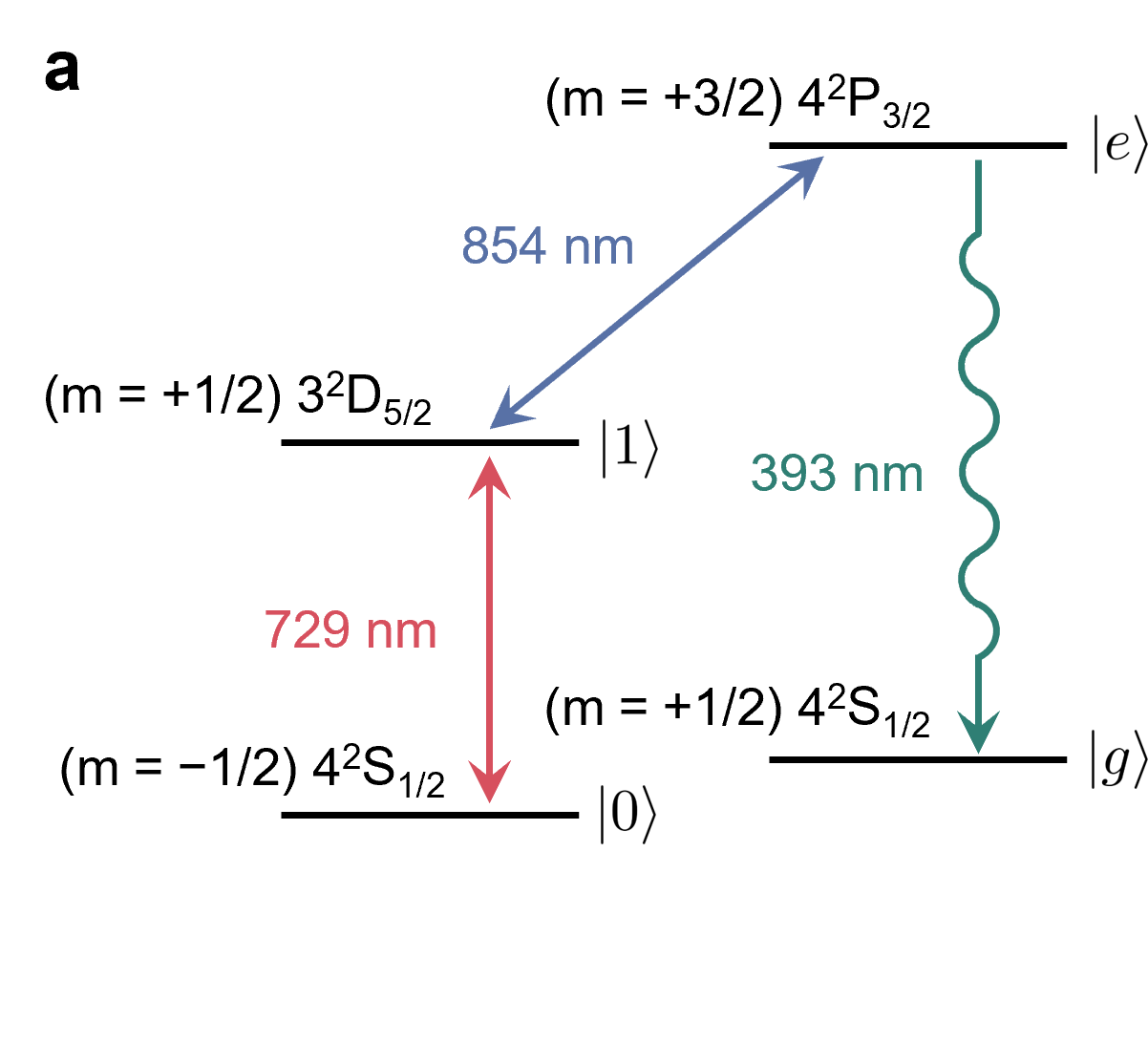}\quad
\includegraphics[width=0.54\textwidth]{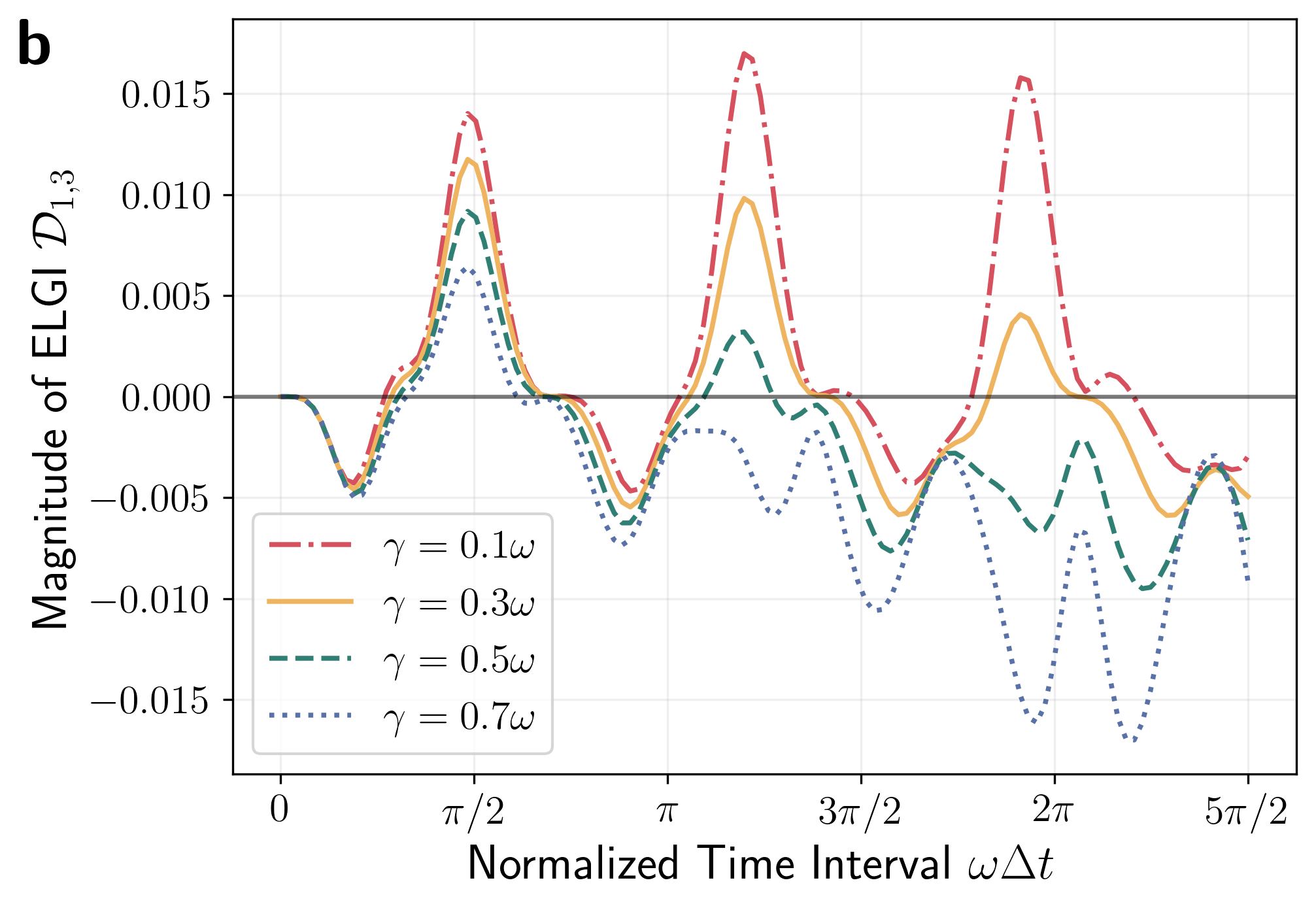}
\caption{{\bfseries Probing non-Markovianity via ELGI violations.}
(a) Schematic of the experimental setup for a trapped $^{40}$Ca$^{+}$ qubit with tunable dissipation \cite{WZZ+21}. The qubit subspace is formed by Zeeman levels $\ket{0}=\ket{4^{2}S_{1/2},m=-1/2}$, $\ket{1}=\ket{3^{2}D_{5/2},m=+1/2}$, and the sink state $\ket{g}=\ket{4^{2}S_{1/2},m=+1/2}$. The $|0\rangle\leftrightarrow|1\rangle$ transition is driven by a 729-nm laser, while a tunable loss from $|1\rangle$ to $|g\rangle$ is induced via the short-lived intermediate state $|e\rangle$ using an 854-nm laser.
(b) Numerical ELGI parameter $\mathcal{D}_{1,3}$ for the dissipative qubit system at different dissipation rates $\gamma=0.1,0.3,0.5,0.7$ (as shown in legend), with $n=3$ equally spaced measurement times $t_{i+1}-t_{i}=\Delta t$. Positive violations of the classical Markovianity condition are observed, which gradually weaken as $\gamma$ increases. Dynamical equations and simulation details are provided in the Supplemental Material.}
\label{fig:non-Markovian}
\end{figure}

To illustrate the applicability of our approach, we consider a dissipative qubit system with decay to a sink state $\ket{g}$, designed to effectively simulate a non-Hermitian qubit dynamics \cite{DSZ+21,WZZ+21}. In practice, the qubit state $\ket{1}$ decays to $\ket{g}$ via an intermediate state $\ket{e}$, with near-Markovian behavior achieved when the dissipation rate $\gamma$ is much larger than the coupling strength $\omega$ between $\ket{1}$ and $\ket{e}$ ($\gamma/\omega \gg 1$); otherwise, non-Markovian effects become significant. As shown in \cref{fig:non-Markovian}, numerical results reveal clear positive violations of the classical Markov condition for moderate dissipation, which gradually weaken as $\gamma$ increases, in agreement with theoretical expectations. Such analyses rely on the premise that ELGIs exhibit negligible quantum violations under the Markov approximation, demonstrating that ELGIs provide a practical and directly measurable tool for probing environmental memory effects in open quantum systems.

More generally, this approach can be naturally extended to detect higher-order non-Markovianity. Following the framework developed by Taranto {\itshape et al.} \cite{TPM+19}, an $(n+1)$-step classical stochastic process can be decomposed into future $F=\{t_{n},\dots,t_{k}\}$, memory $M=\{t_{k-1},\dots,t_{k-l}\}$, and history $H=\{t_{k-l-1},\dots,t_{0}\}$ components, with the corresponding outcomes denoted by $Q^{(F)}$, $Q^{(M)}$, and $Q^{(H)}$. Markovianity requires the Shannon-type equalities
\begin{equation}
I(Q^{(F)};Q^{(H)}|Q^{(M)})=H(Q^{(F)}|Q^{(M)})-H(Q^{(F)}|Q^{(H)},Q^{(M)})=0,
\end{equation}
each of which can be directly mapped onto a corresponding ELGI. In regimes where the Markovian approximation predicts no violation, any observed positive violation of this equality constitutes a clear signature of multi-step memory effects. We expect that future research will further extend the application of ELGIs to a broader range of scenarios, enabling the characterization of additional dynamical features in quantum systems.

\section{Entropic Leggett--Garg inequalities in large-spin systems}\label{sec:large-spin}

Based on the general framework of entropic Leggett--Garg inequalities, we now turn to the spin-$j$ system and investigate the asymptotic behavior of ELGI violations in the limit $j\to+\infty$, to examine whether quantum properties are inevitably excluded by macrorealism at the macroscopic scale. 

Consider a spin system with Hamiltonian $\hat{H} = \omega \hat{J}_{y}$ and observable $\hat{Q} = \hat{J}_{z}$. In the Heisenberg picture, the projection operator corresponding to outcome $m_{i}$ measured at time $t_{i}$ is given by $\hat{\Pi}_{m_{i}}(t_{i}) = \hat{U}^\dagger\big(t_{i}\big)\ket{j,m_{i}}\bra{j,m_{i}}\hat{U}\big(t_{i}\big)$, where $\hat{U}(t)=\exp(-i\omega\hat{J}_{y}t/\hbar)$ denotes the unitary evolution operator. According to Leggett's experimental protocol, one conducts a series of measurements at time $t_{1},\dots,t_{k}$, resulting in outcomes $m_{1},\dots,m_{k}$ respectively. It is important to note that, within the standard quantum formalism, such sequential projective measurements necessarily disturb the system state, which directly violates the assumption of non-invasive measurability postulated by macrorealism. Quantum mechanics predicts the joint probabilities to be
\begin{align}
p(m_{1},\cdots,m_{k})&=\mathrm{Tr}\,\big(\hat{\Pi}_{m_{k-1}}(t_{k-1})\cdots\hat{\Pi}_{m_{1}}(t_{1})\hat{\rho}_{0}\hat{\Pi}_{m_{1}}(t_{1})\cdots\hat{\Pi}_{m_{k-1}}(t_{k-1})\hat{\Pi}_{m_{k}}(t_{k})\big)\notag\\
&=\bra{j,m_{1}}\hat{\rho}_{0}\ket{j,m_{1}}\prod_{i=1}^{k-1}{\big|d_{m_{i+1}m_{i}}^{j}(\beta_{i,i+1})\big|^{2}},
\end{align}
where $\beta_{ij}=\omega(t_{j}-t_{i})$ denotes the rotation angle about $y$-axis, $\hat{\rho}_{0}$ is the density operator of initial state, and $d_{mn}^{j}(\beta)=\bra{j,n}\exp(-i\beta\hat{J}_{y}/\hbar)\ket{j,m}$ are the Wigner $d$-matrix elements.

Previous works have reported LGI violations in the large-spin limit, with maximal violations observed for pure states \cite{KB08,BE14,MRD+19}, and have demonstrated that violations can also occur for maximally mixed states \cite{KB07,MDH16}. Nevertheless, these investigations concentrate on particular measurement settings and parameter regimes, and scaling behavior of both the magnitude and the range of LGI violations with system size has yet to be addressed. Here, we extend the analysis to arbitrary experimental settings, systematically investigating how the violation scales as the system size increases. Specifically, we consider the case where the initial state is maximally mixed, $\hat{\rho}=\hat{I}/(2j+1)$, ensuring that the system initially possesses no quantum coherence. In this case, quantum mechanical joint entropies are simplified as
\begin{equation}
H\big(Q^{(\alpha)}(t_{i}),\cdots,Q^{(\alpha)}(t_{l})\big)=\ln{(2j+1)}+\sum_{s,s'}{H_{j}(\beta_{s,s'})}\ ,
\end{equation}
where $s$ and $s'$ are adjacent measurements in the $\alpha$-th experiment, and $H_{j}(\beta)$ is the ``entropy'' of the Wigner matrix element, defined as
\begin{equation}\label{eqn:Wigner-entropy}
H_{j}(\beta):=-\frac{1}{2j+1}\sum_{m,n=-j}^{j}{\big|d_{nm}^{j}(\beta)\big|^{2}\ln{\big|d_{nm}^{j}(\beta)\big|^{2}}}\ .
\end{equation}

Based on the above analysis, the expressions of $\mathcal{D}_{i}$ and $\mathcal{D}_{i,k}$ are simplified into
\begin{align}
\mathcal{D}_{i}&=H_{j}(\beta_{i-1,i})+H_{j}(\beta_{i,i+1})-H_{j}(\beta_{i-1,i+1})\ ;\\
\mathcal{D}_{i,k}&=\begin{cases}
H_{j}(\beta_{i,i+2})+H_{j}(\beta_{i-1,i+1})-H_{j}(\beta_{i,i+1})-H_{j}(\beta_{i-1,i+2}),&k=i+1\ ;\\
0,&k>i+1\ ;
\end{cases}\label{eqn:mutual-information-simplified}
\end{align}
where the conditions $i\ne1,n$ and $k\ne n$ hold. For boundary cases, the expression reads
\begin{align}
\mathcal{D}_{i}&=\begin{cases}
H_{j}(\beta_{1,2}),&i=1\ ;\\
H_{j}(\beta_{n-1,n}),&i=n\ .
\end{cases}\\
\mathcal{D}_{i,i+1}&=\begin{cases}
H_{j}(\beta_{1,3})-H_{j}(\beta_{1,2}),&i=1,n\ne2\ ;\\
H_{j}(\beta_{n-2,n})-H_{j}(\beta_{n-1,n}),&i=n-1,n\ne2\ ;\\
-H_{j}(\beta_{1,2}),&i=1,n=2\ .
\end{cases}
\end{align}
As will be shown in \cref{eqn:positive-WKB}, in the macroscopic limit $j\to\infty$, $\mathcal{D}_{i}\to+\infty$, and thus remains non-negative. Consequently, only ELGIs of the form $\mathcal{D}_{i,i+1}\ge0$ are capable of detecting macroscopic quantum violations.

Rigorous calculation of ELGIs in spin system involves inevitable computation of $H_{j}(\beta)$, which raises significant challenges to both numerical and analytical treatments, due to the mathematical complexity of the Wigner $d$-matrix elements $d_{mn}^{j}(\beta)$ \cite{FWYJ15,Hof18,WGWS22}. Instead, we then employ the WKB method \cite{BM72} for sufficiently large $j$. It is noteworthy that the spin system under consideration admits the symmetric top as a classical analogue \cite{Wig59}, with the  Schr\"odinger equation \cite{KR27} given by
\begin{equation}
\left(\frac{\mathrm{d}^{2}}{\mathrm{d}\beta^{2}}+\frac{\cos{\beta}}{\sin{\beta}}\frac{\mathrm{d}} {\mathrm{d}\beta}+j(j+1)-\frac{m^{2}-2mn\cos{\beta}+n^{2}}{\sin^{2}{\beta}}\right)d_{mn}^{j}(\beta)=0\ .
\end{equation}
Here $m$ and $n$ denote the angular momentum components along the rotation axis and the $z$-axis, respectively; $\beta$ represents the angle between these two axes; and $j$ is the total angular momentum. For most parameter settings, we obtain the asymptotic expression
\begin{equation}\label{eqn:semi-approx}
H_{j}(\beta)=\ln{(2j+1)\pi}+\ln{|\sin{\beta}|}-\frac{5}{2}+o(1)\ .
\end{equation}
Note that the above asymptotic expression diverges as $\beta\to k\pi\,(k\in\mathbb{Z})$, indicating the semiclassical approximation tends to be invalid, which will be addressed in next section. Substituting $H_{j}(\beta)$ into \cref{eqn:mutual-information-simplified}, we obtain
\begin{align}
\mathcal{D}_{i}&=\ln{(2j+1)\pi}+\ln{\left|1+\tan{\beta_{i-1,i}}\tan{\beta_{i,i+1}}\right|}-\frac{5}{2}+o(1)\ ,\label{eqn:positive-WKB}\\
\mathcal{D}_{i,i+1}&=\ln{\left|\frac{\sin{\beta_{i-1,i+1}}\sin{\beta_{i,i+2}}}{\sin{\beta_{i-1,i+2}}\sin{\beta_{i,i+1}}}\right|} + o(1)\ ,\label{eqn:ELGI-violation}
\end{align}
which demonstrates that the maximal violation of the ELGI is of order $O(1)$. More precisely, as $j \to +\infty$, the absolute magnitude of ELGI violation converges to a value determined solely by the measurement time intervals, as specified by the rotation angles $\beta_{i,i+1}$. However, since the von Neumann entropy $S(\hat{\rho})$ of the system grows logarithmically with $j$ as $S(\hat{\rho})=\ln{(2j+1)}$, the relative violation, expressed as the ratio of the violation magnitude to the system entropy, declines logarithmically to zero as $j$ increases.

\begin{figure}[ht]
\centering
\includegraphics[width=\textwidth]{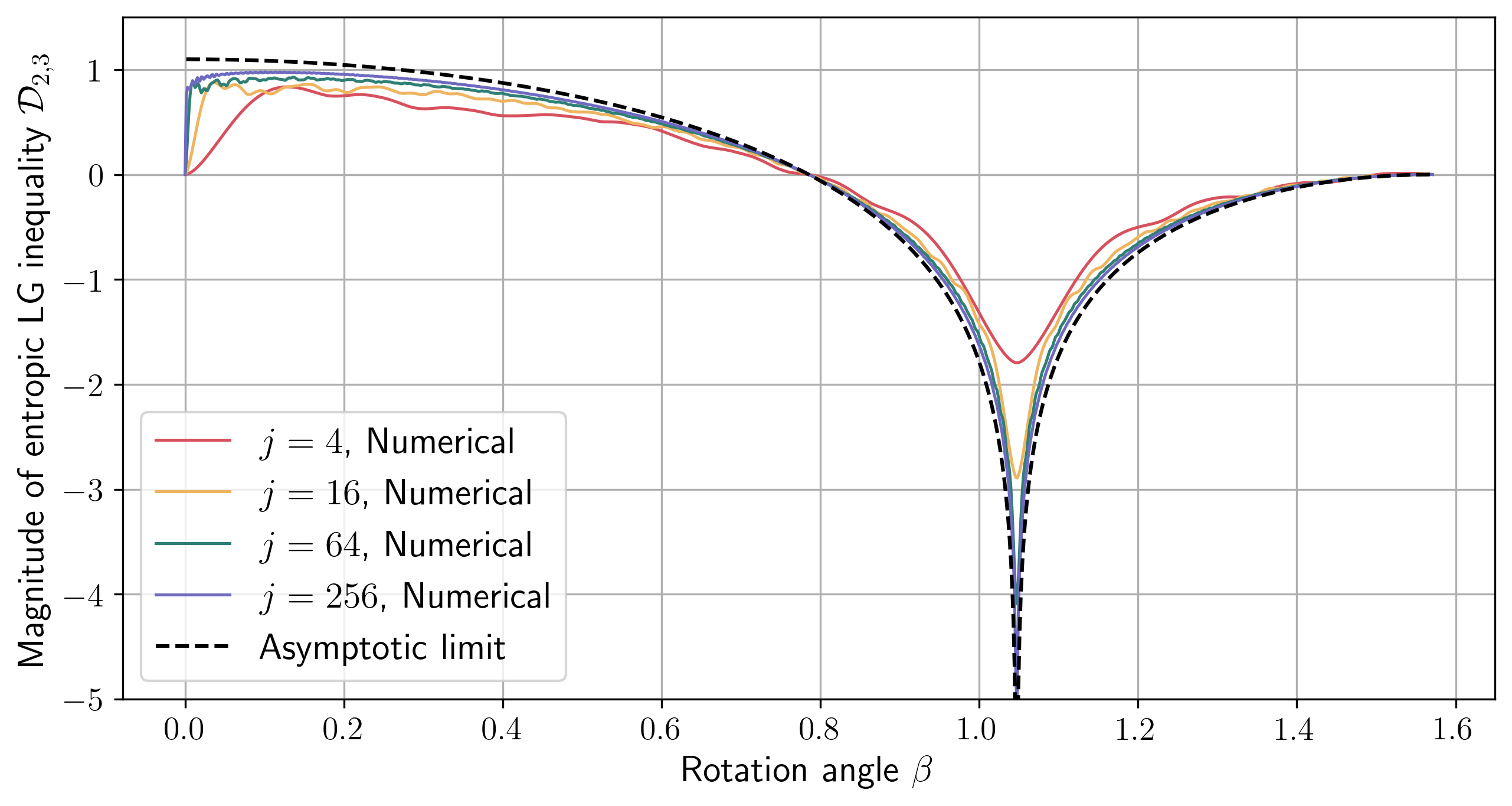}
\caption{{\bfseries Comparison of numerical results with asymptotic limits.} $n=3$ measurement time points are considered, and the rotation angles are set as $\beta_{1,2}=2\beta$ and $\beta_{2,3}=\beta$, yielding an asymptotic limit of $\mathcal{D}_{2,3}=-\ln{|\sin{3\beta}/\sin{\beta}|}$. Because of the symmetry of the Wigner $d$-matrix, calculations are restricted to $0\le\beta\le\pi$ without loss of generality. Apart from the divergence at $\beta=\pi/3$, where $\beta_{1,3}\to\pi$, the numerical results closely match the asymptotic limit as the spin $j$ increases.}\label{fig:WKB-approx}
\end{figure}

Although the derivation of \cref{eqn:ELGI-violation} employs the Shannon cone approximation, considering that distance between the boundaries of $\Gamma_{\text{Sh}}$ and $\Gamma_{\text{E}}$ is a constant independent of $j$, the discussion above can be extended to any entropic Leggett--Garg inequalities, not only the Shannon-type ones. From experimental perspective, quantum violation tends progressively to be more challenging to detect at larger scales. We observe that the magnitude of the violation remains bounded, while the background entropy expands logarithmically. This scaling behavior implies that the violation signal is effectively submerged by the macroscopic uncertainty, like searching for tiny shells in a rising tide. Notably, this result suggests a complementary mechanism for the emergence of macrorealism. Unlike the coarse-graining restriction proposed by Kofler and Brukner \cite{KB07,KB08}, which attributes classicality to limited measurement resolution, our findings indicate that even under ideal sharp measurements, quantum features can become operationally negligible due to intrinsic information-theoretic scaling constraints.

\section{Macroscopic quantum violations beyond the WKB approximation}

Although the numerical results generally agree with the asymptotic limit across most parameter ranges, as illustrated in \cref{fig:WKB-approx}, there are exceptions identified in \cref{eqn:ELGI-violation} exhibiting trouble divergences. Kofler and Brukner \cite{KB07} provided a relevant example, showing that quantum violation of the maximally mixed state near $\beta=0$ persist in the $j\to+\infty$ limit. We further demonstrate that such a phenomenon applies to the discrete sequence $\beta = n\pi\,(n\in\mathbb{Z})$, arising from the failure of the WKB approximation.

\begin{figure}[ht]
\centering
\includegraphics[width=0.95\textwidth]{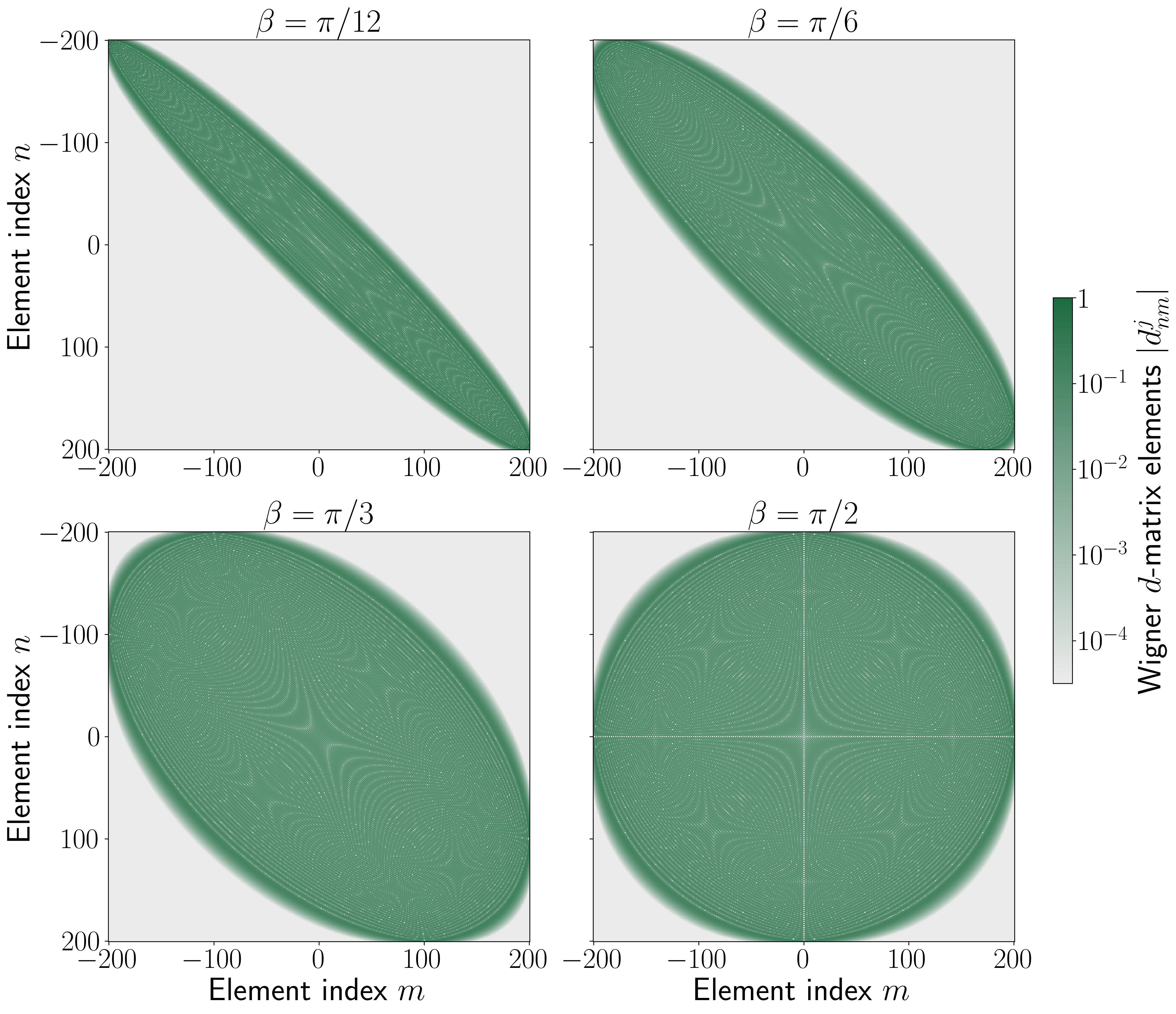}
\caption{{\bfseries Numerical calculation of Wigner $\bm{d}$-matrix elements.} Wigner $d$-matrix elements $d_{mn}^{j}(\beta)$ are calculated at rotation angles $\beta=\pi/12,\pi/6,\pi/3,\pi/2$ for $j=200$, and plotted on semi-logarithmic axes. The elliptical region with rapid oscillations corresponds to the classically allowed region, while outside the boundary the matrix elements exhibit exponential decay in classically forbidden region. As $\beta$ decreases from $\pi/2$ to zero, the ellipse’s eccentricity increases until it degenerates into a line at $\beta=0$ (not shown).}\label{fig:Wigner-elem}
\end{figure}

In preceding section, the Schr\"odinger equation is handled by means of the WKB approximation, with the solution assumed to take the form of $d_{mn}^{j}(\beta) \approx Ae^{iS(\beta)}$. Here, $S(\beta)$ denotes the action associated with the nutation angle $\beta$, i.e.
\begin{equation}\label{eqn:quantum-action}
S(\beta)=\int_{\beta_{0}}^{\beta}{\frac{\sqrt{J^{2}\sin^{2}{\alpha}-m^{2}-n^{2}+2mn\cos{\alpha}}}{J\sin{\alpha}}\,\mathrm{d}\alpha}\ ,
\end{equation}
where $\beta_{0}$ denotes the zero of the integrand, and $J=j+1/2$ for simplicity. The Schr\"odinger equation reduces to the Hamilton--Jacobi equation in the limit $\hbar\to0$, causing the quantum action to coincide with its classical analogue. According to \cref{eqn:quantum-action}, however, the quantum action can be real or purely imaginary, depending on the sign of the discriminant $R(m,n,\beta)=J^{2}\sin^{2}{\beta}-m^{2}-n^{2}+2mn\cos{\beta}$, whereas the classical action must be real. Imaginary action corresponds to a trajectory in phase space denied by classical mechanics, as the canonical variables are complex \cite{BT57,BGHS96,BL98}. Hence, for a fixed rotation angle $\beta$, the Wigner $d$-matrix elements $d_{mn}^{j}(\beta)$ can be divided into two regions, as shown in \cref{fig:Wigner-elem}: a ``classically allowed'' region with real action, characterized by rapid oscillations; and a ``classically forbidden'' region with imaginary action, where the matrix elements exponentially decays.

Though the WKB approximation proves accurate in both domains, it diverges near the classical-quantum boundary, which, for most parameter settings, proves to be inconsequential in \cref{sec:large-spin}. For instance, in the calculation of $H_{j}(\beta)$, analysis indicates that the semi-classical approximation fails within a boundary layer of width $\delta\sim j^{-1/3}$, where the number of matrix elements is an infinitesimal compared to the classically allowed region, and hence the WKB approximation is legitimate to apply to the leading-order asymptotic terms.

Nevertheless, when $\beta\to n\pi\,(n\in\mathbb{Z})$, the classically allowed region degenerates from an ellipse to a line, causing the WKB approximation fail entirely within the neighborhood of width $\Delta\beta\sim j^{-1}$. In this regime, we obtain
\begin{equation}
\mathcal{D}_{i,i+1}\sim\begin{cases}
-\ln{(2j+1)}+O(\beta^{2}),&\beta_{i,i+1}\to n\pi\ ;\\
\ln{(2j+1)}+O(\beta^{2}),&\beta_{i-1,i}\to n\pi\ \text{or}\ \beta_{i+1,i+2}\to n\pi;
\end{cases}
\end{equation}
which indicates when $\beta_{i,i+1}\to n\pi\,(n\in\mathbb{Z})$, the quantum violation approaches to maximum rapidly. This result provides an experimental evidence of contradiction between quantum mechanics and the classical worldview. Remarkably, the magnitude of the quantum violation increases with the scale of the system, while the quantum resources required do not scale accordingly, as the conclusion is drawn from maximally mixed state.

This counterintuitive conclusion arises from the simple fact that, the ensemble average of the time correlation is not equal to the time correlation of the ensemble average. Consequently, the correlations observed from sequential measurements provide a more comprehensive characterization of quantum phenomena, which is inaccessible per analysis of the dynamical evolution of density matrix alone.

\section{Conclusion and outlook}

In this study, we established a systematic geometric framework for entropic Leggett--Garg inequalities, providing a unified methodology for deriving inequalities that include higher-order temporal correlations. Numerical results demonstrate that ELGIs exhibit quantum violations across broader parameter regimes than conventional LGIs. ELGIs also exhibit enhanced robustness against decoherence, maintaining observable violations under stronger dissipation and over extended time intervals. These characteristics position ELGIs as a robust, information-theoretically grounded tool for certifying quantum coherence in noisy or open quantum systems. Notably, the information-theoretic foundation of our framework provides a transparent physical interpretation, enabling the operational distinction between measurement invasiveness and environmental memory effects in open quantum systems.

The WKB analysis in large-spin limit reveals that, for an initially maximally mixed system, quantum violation occurs exclusively in ELGIs of the form $\mathcal{D}_{i,i+1}\ge0$. In generic parameter regimes, the magnitude of these violations saturates at a finite value, making them increasingly negligible relative to the logarithmically growing system entropy. Nevertheless, in the degenerate case where the rotation angle $\beta_{i,i+1}\to k\pi$, the quantum violation rapidly approaches the maximal value $\mathcal{D}_{i,i+1}\to-\ln{(2j+1)}$. This observation provides a clear pathway for quantum violation detection in large-scale system and for the detection of fundamental boundaries of quantum mechanics.

These findings offer a complementary perspective on the emergence of classicality at the macroscopic scale. While the fundamental incompatibility between quantum and classical descriptions persists at the macroscopic scale, the prevalence of classical behavior implies a robust suppression mechanism for quantum effects. Unlike the coarse-graining restriction \cite{KB07}, which attributes this transition to limited measurement resolution, our results demonstrate that even under ideal sharp measurements, quantum violations remain effectively negligible in most parameter regimes due to intrinsic information-theoretic scaling constraints. Consequently, macroscopic quantum phenomena do not vanish but persist only in specific, fine-tuned regimes. Our results provide a clear pathway for detecting quantum violations in large-scale systems and for probing the fundamental boundaries of quantum mechanics.

Finally, since the analysis is limited to unitary evolution, a natural question arises that whether macroscopic quantum violations can still persist in open system subject to dissipation or not. If not, how do such violations diminish as the strength of dissipation increases? Further investigation of this open question will advance our understanding of the decoherence mechanism, by which quantum systems become effectively classical at macroscopic scales, and will clarify the potential of the ELGI framework.

\backmatter

{\bmhead{Acknowledgements}

This work was supported in part by National Natural Science Foundation of China(NSFC) under the Grants 12475087 and 12235008, and University of Chinese Academy of Sciences.}

{\bmhead{Author Contributions}

C.F.Q. conceived and supervised the research. Q.H.C., X.H.Y., M.C.Y., A.X.L., and C.F.Q. performed the theoretical analysis. X.H.Y., Q.H.C., M.C.Y., A.X.L., and C.F.Q. performed the numerical calculation. All authors contributed to discussing, editing, and revising the manuscript.}

{\bmhead{Competing interests}

All authors declare no financial or non-financial competing interests. }

{\bmhead{Data Availability}

The code and numerical data supporting the findings of this study are openly available on GitHub at: https://github.com/PolarisMegrez/Entropic-Leggett-Garg-Inequalities. No additional datasets were generated or analyzed beyond the materials provided in this repository.}

\appendix
\renewcommand{\theequation}{S\arabic{equation}}
\setcounter{equation}{0}
\newgeometry{left=2cm,right=2cm,top=2.5cm,bottom=2.5cm}

\renewcommand{\thefigure}{S\arabic{figure}}

\section{Derivation of the entropic Leggett--Garg inequalities in fig. 2}

In the main text, we demonstrate the capacity of entropic Leggett-Garg inequalities (ELGIs) of varying orders to reveal quantum violations, as illustrated in fig. 2. The order of an ELGI corresponds to the number of variables involved in its joint entropy terms; for instance, $H(Q_{1}, Q_{2})$ denotes a second-order entropy, while $H(Q_{1}, Q_{2}, Q_{3})$ denotes a third-order entropy. In this section, we illustrate the procedure for systematically constructing inequalities of arbitrary order based on the elementary-type ELGIs.

For the case of $n=3$, elementary-type ELGI is divided into four groups, namely
\begin{align}
\mathcal{D}_{i}^{(123)}&=H(Q_{1},Q_{2},Q_{3})-H(Q_{m},Q_{n})\ge0\ ,\quad(m,n\ne i)\label{eqn:third-order-a}\\
\mathcal{D}_{ij}^{(123)}&=H(Q_{i},Q_{m})+H(Q_{j},Q_{m})-H(Q_{1},Q_{2},Q_{3})-H(Q_{m})\ge0\ ,\quad(m\ne i,j)\label{eqn:third-order-b}\\
\mathcal{D}_{k}^{(ij)}&=H(Q_{i},Q_{j})-H(Q_{m})\ge0\ ,\quad(m\ne i,j)\label{eqn:second-order-a}\\
\mathcal{D}_{ij}^{(ij)}&=H(Q_{i})+H(Q_{j})-H(Q_{i},Q_{j})\ge0\ ,\label{eqn:second-order-b}
\end{align}
where $i,j\in\{1,2,3\}$ are distinct indices, and the superscripts indicate available time points.

The above elementary-type ELGIs collectively constitute the necessary and sufficient conditions for all third-order ELGIs. It is important to emphasize that restricting the condition to $D_{i}^{(123)}\ge0$ and $D_{ij}^{(123)}\ge0$ alone is insufficient, as distinct third-order ELGIs can be constructed by combining with second-order ELGIs. For example, we obtain
\begin{equation}
\mathcal{D}_{k}^{(123)}+\mathcal{D}_{k}^{(ij)}=H(Q_{1},Q_{2},Q_{3})-H(Q_{k})\ge0\ .\quad(k=i,j)
\end{equation}

In order to obtain the necessary and sufficient conditions for all second-order ELGIs, we need to remove the third-order term $H(Q_{1}, Q_{2}, Q_{3})$ from the elementary-type ELGIs. In principle, the elimination of specified variables from a system of linear inequalities can be systematically performed via the Fourier-Motzkin elimination method \cite{Wil86}. For instance, in the case of $n=3$, we ultimately obtain all possible combinations of $\mathcal{D}_{k}^{(123)}+\mathcal{D}_{i,j}^{(123)}$, namely,
\begin{equation}
\mathcal{D}^{(123)}_{i,j|k}:=\mathcal{D}_{k}^{(123)}+\mathcal{D}_{i,j}^{(123)}=H(Q_{i},Q_{k})+H(Q_{j},Q_{k})-H(Q_{k})-H(Q_{i,j})\ge0\ .\label{eqn:second-order-c}
\end{equation}

For the case $n=4$, the elementary-type ELGI is divided into six groups, namely
\begin{align}
\mathcal{D}_{i}^{(1234)}&=H(Q_{1},Q_{2},Q_{3},Q_{4})-H(Q_{r},Q_{s},Q_{t})\ge0\ ,\quad(r,s,t\ne i)\label{eqn:fourth-order-a}\\
\mathcal{D}_{ij}^{(1234)}&=H(Q_{i},Q_{r},Q_{s})+H(Q_{j},Q_{r},Q_{s})-H(Q_{1},Q_{2},Q_{3},Q_{4})-H(Q_{r},Q_{s})\ge0\ ,\quad(r,s\ne i,j)\label{eqn:fourth-order-b}\\
\mathcal{D}_{m}^{(ijk)}&=H(Q_{i},Q_{j},Q_{k})-H(Q_{r},Q_{s})\ge0\ ,\quad(r,s=i,j,k; r,s\ne m)\label{eqn:third-order-c}\\
\mathcal{D}_{mn}^{(ijk)}&=H(Q_{m},Q_{r})+H(Q_{n},Q_{r})-H(Q_{i},Q_{j},Q_{k})-H(Q_{m},Q_{n})\ge0\ ,\quad(m,n,r=i,j,k; r\ne m,n)\label{eqn:third-order-d}\\
\mathcal{D}_{m}^{(ij)}&=H(Q_{i},Q_{j})-H(Q_{r})\ge0\ ,\quad(m,r=i,j; r\ne i,j)\label{eqn:second-order-d}\\
\mathcal{D}_{ij}^{(ij)}&=H(Q_{i})+H(Q_{j})-H(Q_{i},Q_{j})\ge0\ ,\label{eqn:second-order-e}
\end{align}
where $i,j,k\in\{1,2,3,4\}$ are distinct indices, and the superscripts indicate available time points.

Besides, after Fourier-Motzkin elimination, we obtain
\begin{align}
\mathcal{D}^{(1234)}_{i,j|k}&:=\mathcal{D}_{k}^{(1234)}+\mathcal{D}_{i,j}^{(1234)}=H(Q_{i},Q_{k},Q_{r})+H(Q_{j},Q_{k},Q_{r})-H(Q_{k},Q_{r})-H(Q_{i},Q_{j},Q_{r})\ge0\ ,\quad(r\ne i,j,k)\label{eqn:third-order-e}\\
\mathcal{D}^{(ijk)}_{m,n|l}&:=\mathcal{D}_{l}^{(ijk)}+\mathcal{D}_{m,n}^{(ijk)}=H(Q_{m},Q_{l})+H(Q_{n},Q_{l})-H(Q_{l})-H(Q_{m,n})\ge0\ .\quad(l,m,n\ne i,j,k)\label{eqn:second-order-f}
\end{align}

Therefore, in fig. 2a, the second-order ELGI contains the inequalities \cref*{eqn:second-order-a,eqn:second-order-b,eqn:second-order-c}, and the third-order ELGI contains the inequalities \cref*{eqn:second-order-a,eqn:second-order-b,eqn:third-order-a,eqn:third-order-b}; in fig. 2b, the second-order ELGI contains the inequalities \cref*{eqn:second-order-c,eqn:second-order-d,eqn:second-order-f}, the third-order ELGI contains the inequalities \cref*{eqn:second-order-d,eqn:second-order-e,eqn:third-order-c,eqn:third-order-d,eqn:third-order-e}, and the four-order ELGI contains the inequalities \cref*{eqn:second-order-d,eqn:second-order-e,eqn:third-order-c,eqn:third-order-d,eqn:fourth-order-a,eqn:fourth-order-b}.

In addition, for the case of $n=3$, the ELGI formulated by Devi {\itshape et al.} is given as
\begin{equation}
H(Q_{1},Q_{2})+H(Q_{2},Q_{3})-H(Q_{2})-H(Q_{1},Q_{3})\ge0\ ;
\end{equation}
whereas for $n=4$, the corresponding ELGIs are
\begin{align}
H(Q_{1},Q_{2})+H(Q_{2},Q_{3})+H(Q_{3},Q_{4})-H(Q_{2})-H(Q_{3})-H(Q_{1},Q_{4})&\ge0\ ,\\
H(Q_{i},Q_{j})+H(Q_{j},Q_{k})-H(Q_{j})-H(Q_{i},Q_{k})&\ge0\ ,
\end{align}
where $i,j,k\in\{1,2,3,4\}$ and $i<j<k$.



\section{Non-Markovian dynamics in a dissipative qubit system}

In this section, we provide the background and model of the trapped dissipative qubit system used to detect non-Markovian effects via ELGIs. The system encodes a qubit in the Zeeman sublevels of the electronic states:
\begin{equation}
\begin{array}{l l}
\ket{0}=\ket{4^{2}S_{1/2},m_{J}=-1/2},\quad&\ket{1}=\ket{3^{2}D_{5/2},m_{J}=+1/2},\\
\ket{g}=\ket{4^{2}S_{1/2},m_{J}=+1/2},\quad&\ket{e}=\ket{4^{2}P_{3/2},m_{J}=+3/2},\\
\end{array}
\end{equation}
where $\ket{g}$ serves as a tunable sink state. As proposed by Wang {\itshape et al.} \cite{WZZ+21}, the $\ket{0}\leftrightarrow\ket{1}$ transition is coherently driven by a 729-nm laser, while a controllable dissipation channel from $\ket{1}$ to $\ket{g}$ is implemented via a short-lived intermediate state $\ket{e}$ using an 854-nm laser.

The full dynamics of this four-level system is described by the Lindblad master equation
\begin{equation}
\frac{\text{d}\hat{\rho}}{\text{d}t}=-i[\hat{H},\hat{\rho}] + \hat{L}\hat{\rho}\hat{L}^\dagger - \frac{1}{2}\{\hat{L}^\dagger\hat{L},\hat{\rho}\},
\end{equation}
where the Hamiltonian takes the form
\begin{equation}
\hat{H}=E_{0}\ket{0}\bra{0}+E_{1}\ket{1}\bra{1}+E_{e}\ket{e}\bra{e}+E_{g}\ket{g}\bra{g}+\omega_{0}(\ket{0}\bra{1}+\ket{1}\bra{0})+\omega(\ket{1}\bra{e}+\ket{e}\bra{1}),
\end{equation}
and the dissipative process is captured by the Lindblad jump operator
\begin{equation}
\hat{L}=\sqrt{\gamma}\,\ket{g}\bra{e}.
\end{equation}

In the regime where the intermediate state $\ket{e}$ decays much faster than the coherent couplings ($\gamma \gg\omega_{0},\omega$), the population of $\ket{e}$ can be adiabatically eliminated, giving rise to an effective three-level model. The Hamiltonian restricted to the qubit subspace is
\begin{equation}
\hat{H}_{\text{eff}}=E_{0}\ket{0}\bra{0}+E_{1}\ket{1}\bra{1}+\omega_{0}(\ket{0}\bra{1}+\ket{1}\bra{0}),
\end{equation}
and the effective decay from $\ket{1}$ to the sink state $\ket{g}$ is described by
\begin{equation}
\hat{L}_{\text{eff}}=\sqrt{\Gamma_{\text{eff}}}\,\ket{g}\bra{1}, \quad 
\Gamma_{\text{eff}}=\frac{|\omega|^2\gamma}{\Delta^2 +(\gamma/2)^2}, \quad
(\Delta=E_{e}-E_{1}).
\end{equation}

While the effective model provides an accurate description of the qubit dynamics in the fast-decay (Markovian) limit, deviations arise when the decay of $\ket{e}$ is slow. In such cases, the full four-level dynamics exhibits memory effects that cannot be fully captured by the simple Markovian effective model, leading to observable differences in temporal correlations and positive violations to classical Markovianity conditions.

Our experiment considers an initial state $\hat{\rho}=(\ket{0}\bra{0}+\ket{1}\bra{1})/2$, which corresponds to the maximally mixed state in the qubit subspace. The first measurement is performed at $\omega t_{1}=\pi/4$, which maximizes the expected violation. We adopt $n=3$ equally spaced measurement times, with $t_{i+1}-t_{i}=\Delta t$, to evaluate the temporal correlations.

Numerical results indicate that for relatively low decay rates ($\gamma < 0.1 \, \omega$), the system exhibits very small quantum (negative) violations of the ELGI under the Markov approximation, $|\mathcal{D}_{1,3}| \sim 10^{-4}$. In contrast, the non-Markovian dynamics of the full four-level system produces a significantly larger violation, $|\mathcal{D}_{1,3}| \sim 10^{-2}$. These results demonstrate that in this regime, where the quantum violation under the Markov approximation is negligible compared to the non-Markovian violation, the protocol can effectively detect and characterize the non-Markovian nature of the system dynamics.

\section{WKB Approximation Method for Deriving Wigner {\itshape \MakeLowercase{d}\,}-Matrix}\label{sec:WKB-solution}

In this section, we briefly review the Wenzel--Kramers--Brillouin approximation method \cite{Bri26,Kra26,Wen26} (also known as the Liouville--Green method \cite{Lio37,Gre37}, especially in mathematical literatures) for approximately deriving the Wigner $d$-matrix elements. For better handling the behaviour of functions $d^{i}_{nm}(\beta)$ near the classical-quantum boundary, we are to follow the so-called {\itshape uniform} approximation method first established by Cherry \cite{Che50} and further discussed by Jeffreys \cite{Jef53}, Erd\'elyi \cite{Erd60} and others, with slightly difference from the original WKB method.

\subsection{Second-order ordinary differential equation for Wigner $d$-matrix}

The WKB method is generally used to find approximate solutions to second-order ordinary differential equations (in physics, usually the Schr\"odinger equation). Wigner \cite[Chap.\,19]{Wig59b} noticed that the functions $D_{nm}^{j}(\alpha,\beta,\gamma)$ are the eigenfunctions of the symmetric top, and thus satisfy the Schr\"odinger equation \cite{Rei26,KR27}
\begin{equation}
\left(\frac{\partial^{2}}{\partial\beta^{2}}+\frac{\cos{\beta}}{\sin{\beta}}\frac{\partial}{\partial\beta}+\big(\frac{I_{x}}{I_{z}}+\frac{\cos^{2}{\beta}}{\sin^{2}{\beta}}\big)\frac{\partial^{2}}{\partial\gamma^{2}}+\frac{1}{\sin^{2}{\beta}}\frac{\partial^{2}}{\partial\alpha^{2}}-\frac{2\cos{\theta}}{\sin^{2}{\theta}}\frac{\partial^{2}}{\partial\alpha\partial\gamma}+\frac{2I_{x}E}{\hbar^{2}}\right)D_{nm}^{j}(\alpha,\beta,\gamma)=0\ ,
\end{equation}
where $I_{x}$ and $I_{z}$ denote for the moments of inertia of the top about the $x$- and $z$-axes (the axis of symmetry), respectively. Given that $D_{nm}^{j}(\alpha,\beta,\gamma)=e^{-im\alpha}d^{j}_{nm}(\beta)e^{-in\gamma}$, we obtain the differential equation satisfied by the Wigner $d$-matrix elements,
\begin{equation}
\left(\frac{\mathrm{d}^{2}}{\mathrm{d}\beta^{2}}+\frac{\cos{\beta}}{\sin{\beta}}\frac{\mathrm{d}}{\mathrm{d}\beta}+j(j+1)-\frac{m^{2}-2mn\cos{\beta}+n^{2}}{\sin^{2}{\beta}}\right)d_{nm}^{j}(\beta)=0\ .
\end{equation}
where the eigenvalues of energy $E$ obtained in Ref.\,\cite{Rei26,KR27} have been substituted: 
\begin{equation}
E_{j,n}=\frac{\hbar^{2}}{2I_{x}}\left(j(j+1)+\big(\frac{I_{x}}{I_{z}}-1\big)n^{2}\right).
\end{equation}

For simplicity, we denote $J=j+1/2,\,\mu=m/J,\,\nu=n/J$ and assume $0<\beta<\pi$. The other elements of Wigner $d$-matrix can be derived with the following symmetry:
\begin{equation}\label{eqn:symmetry}
d_{m,n}^{j}(-\beta)=d_{n,m}^{j}(\beta)=(-1)^{m-n}d_{m,n}^{j}(\beta)\ .
\end{equation}
Thus, the function $w(\beta)=\sqrt{\sin{\beta}}\cdot d^{j}_{nm}(\beta)$ obeys the differential equation
\begin{equation}\label{eqn:original-eqn}
\left(\frac{\mathrm{d}^{2}}{\mathrm{d}\beta^{2}}+\frac{J^{2}}{\sin^{2}{\beta}}\big(\sin^{2}{\beta}-\mu^{2}-\nu^{2}+2\mu\nu\cos{\beta}\big)+\frac{1}{4\sin^{2}{\beta}}\right)w(\beta)=0\ .
\end{equation}

\subsection{Uniform WKB method}

Now we start to deal with \cref{eqn:original-eqn} following the {\itshape uniform} WKB method. The core of this method is to introduce an ancilla parameter $\zeta$, which satisfies equation
\begin{equation}\label{eqn:zeta-eqn}
\zeta\big(\frac{\mathrm{d}\zeta}{\mathrm{d}\beta}\big)^{2}=-\frac{J^{2}}{\sin^{2}{\beta}}R(\beta)\ ,\qquad\left( R(\beta)=\sin^{2}{\beta}-\mu^{2}-\nu^{2}+2\mu\nu\cos{\beta}\,\right)
\end{equation}
and the corresponding ancilla function
\begin{equation}
h(\zeta)=\left|\frac{\mathrm{d}\zeta}{\mathrm{d}\beta}\right|^{1/2}\cdot w(\beta)=\sqrt{\frac{J}{\sin{\beta}}}\Big(\frac{R(\beta)}{-\zeta}\Big)^{1/4}\cdot w(\beta)\ ,
\end{equation}
Here, $R(\beta)$ is the discriminant of \cref{eqn:original-eqn}, with its zeros $\beta_{\pm}=\arccos{\big(\mu\nu\pm\sqrt{(1-\mu^{2})(1-\nu^{2})}\,\big)}$ being the two turning points of the differential equation. 

Our intention is to change the variable of \cref{eqn:original-eqn}, thereby obtaining the {\itshape standard} form of the second-order differential equation, (cf.\,\cite[Lem.\,1]{Che50})
\begin{equation}\label{eqn:Airy-eqn}
\frac{\mathrm{d}^{2}h}{\mathrm{d}\zeta^{2}}=\left(\zeta+\left|\frac{\mathrm{d}\beta}{\mathrm{d}\zeta}\right|^{1/2}\frac{\mathrm{d}^{2}}{\mathrm{d}\zeta^{2}}\left|\frac{\mathrm{d}\zeta}{\mathrm{d}\beta}\right|^{1/2}+\frac{\zeta}{4J^{2}R(\beta)}\right)h(\zeta)=\left(1+\frac{1}{J^{2}}\cdot\frac{\delta(\beta)}{4R^{3}(\beta)}\right)\zeta\cdot h(\zeta)\ ,
\end{equation}
where
\begin{equation}
\delta(\beta)=\frac{5J^{2}R^{3}(\beta)}{4\zeta^{3}(\beta)}+\big(1-\mu^{4}-\nu^{4}+3\mu^{2}\nu^{2}-6\mu\nu\cos{\beta}-2\mu\nu\cos^{3}{\beta}-(1-4\mu^{2}-4\nu^{2}+\mu^{2}\nu^{2})\cos^{2}{\beta}\big)\sin^{2}{\beta}\ ,
\end{equation}
is bounded and independent of $J$, as can be seen from \cref{eqn:action-func,eqn:zeta-exp} below. Thus, for $J$ sufficiently large, we can approximately obtain
\begin{equation}\label{eqn:airy}
\frac{\mathrm{d}^{2}h}{\mathrm{d}\zeta^{2}}=\zeta\cdot h(\zeta)\quad\to\quad h(\zeta)=C(\mu,\nu)\cdot\Ai(\zeta)\ ,
\end{equation}
where $\Ai(\zeta)$ represents for the Airy function, the coefficient $C(\mu,\nu)$ is determined through the initial conditions, and we rule out the other solutions with $\Bi(\zeta)$ components, since $h(\zeta)$ should be bounded.

Clearly, $\zeta$ is required to have an one-to-one correspondence with $\beta$, and $\zeta'(\beta)\ne0$ for any $\beta\in(0,\pi)$, which is actually impossible as $\zeta(\beta_{\pm})=0$ according to \cref{eqn:zeta-eqn}. Neverthelss, by allowing $\zeta(\beta)$ to take the value of zero at only one of the two turning points $\beta_{\pm}$, we obtain a valid solution $\zeta(\beta)$ covering the interval $(0,\beta_{+})$ or $(\beta_{-},\pi)$, respectively.
\begin{equation}\label{eqn:zeta-exp}
\zeta(\beta)=\begin{cases}
\displaystyle-\left(\frac{3J}{2}S(\beta)\right)^{2/3},&R(\beta)>0\ ;\\[10pt]
\displaystyle\left(-\frac{3J}{2}S(\beta)\right)^{2/3},&R(\beta)<0\ .
\end{cases}
\end{equation}
Here, we have introduced a function $S(\beta)$ for simplicity, corresponding to the classical action of the trajectory connecting the initial and final state (cf. \cite[Sec.\,2]{BGHS96}), namely
\begin{equation}\label{eqn:action-func}
S_{\pm}(\beta)=\begin{cases}
\begin{array}{r}
\displaystyle\pm\arctan{\frac{\cos{\beta}-\mu\nu}{\sqrt{R(\beta)}}}\mp\frac{1}{2}(\mu+\nu)\arctan{\frac{(\cos{\beta}-\mu\nu)(1+\mu\nu)+(1-\mu^{2})(1-\nu^{2})}{(\mu+\nu)\sqrt{R(\beta)}}}\qquad\qquad\quad\ \\[10pt]
\displaystyle\pm\frac{1}{2}(\mu-\nu)\arctan{\frac{(\mu\nu-\cos{\beta})(1-\mu\nu)+(1-\mu^{2})(1-\nu^{2})}{(\mu-\nu)\sqrt{R(\beta)}}}-\big|\mu+\nu\big|\frac{\pi}{4}-\big|\mu-\nu\big|\frac{\pi}{4}+\frac{\pi}{2}\ ,\\
\end{array}&\big(R(\beta)>0\big)\\[30pt]
\begin{array}{r}
\displaystyle\mp\arcoth{\frac{\cos{\beta}-\mu\nu}{\sqrt{-R(\beta)}}}\pm\frac{1}{2}(\mu+\nu)\arcoth{\frac{(\cos{\beta}-\mu\nu)(1+\mu\nu)+(1-\mu^{2})(1-\nu^{2})}{(\mu+\nu)\sqrt{-R(\beta)}}}\qquad\qquad\quad\ \\[10pt]
\displaystyle\mp\frac{1}{2}(\mu-\nu)\arcoth{\frac{(\mu\nu-\cos{\beta})(1-\mu\nu)+(1-\mu^{2})(1-\nu^{2})}{(\mu-\nu)\sqrt{-R(\beta)}}}\ ,\\
\end{array}&\big(R(\beta)<0\big)
\end{cases}
\end{equation}
derived from
\begin{equation}\label{eqn:action-diff}
\frac{\partial S_{\pm}(\beta)}{\partial\beta}=\mp\frac{\sqrt{|R(\beta)|}}{\sin{\beta}},\quad S_{\pm}(\beta_{\pm})=0\ ,
\end{equation}
where the subscript denotes which the turning point we have choose.

After substitution, we obtain the approximate expression
\begin{align}
d^{j}_{nm}(\beta)\approx\frac{C(\mu,\nu)\cdot\Ai(\zeta)}{\sqrt{J}}\Big(\frac{-\zeta}{R(\beta)}\Big)^{1/4}.
\end{align}

\subsection{Boundary conditions of the Wigner $d$-matrix}

Now, we are only one step away from a complete approximate expression for $d_{nm}^{j}(\beta)$, which is to determine the coefficients $C(\mu,\nu)$. Given that for any $\mu,\nu$ and $j$, there exists a sufficiently small $\beta$ to make our approximation valid, we could derive the expression of $C(\mu,\nu)$ by comparing the coefficient of the leading-order term of $d_{nm}^{j}(\beta)$ when $\beta\to0$. We start with Wigner's series expansion for $d_{nm}^{j}(\beta)$ in Ref.\,\cite[Chap.\,15]{Wig59a}, namely
\begin{equation}\label{eqn:Wigner-series}
d^{j}_{nm}(\beta)=\sqrt{(j+m)!(j-m)!(j+n)!(j-n)!}\sum_{s=s_{\min}}^{s_{\max}}{\frac{(-1)^{n-m+s}\big(\cos{(\beta/2)}\big)^{2j+m-n-2s}\big(\sin{(\beta/2)}\big)^{n-m+2s}}{(j+m-s)!s!(n-m+s)!(j-n-s)!}}\ ,
\end{equation}
where $s_{\min}=\max\{0,\,m-n\}$ and $s_{\max}=\min\{j+m,\,j-n\}$, thus we obtain
\begin{equation}
d^{j}_{nm}(\beta)=\sqrt{\frac{(j+a)!(j-b)!}{(j+b)!(j-a)!}}\frac{(-1)^{a-b}}{(a-b)!}\cdot\big(\frac{\beta}{2}\big)^{a-b}+O(\beta^{a-b+2})\ .
\end{equation}
where $a=\max(m,n),b=\min(m,n)$.

Given that $S(\beta)\sim O(\ln{\beta})$ as $\beta\to0$ as can be seen from \cref{eqn:action-diff}, we use the asymptotic formulae for Airy function as the variable tends to be infinity:
\begin{equation}\label{eqn:Airy-asymptotic}
\Ai(x)=\begin{cases}
\displaystyle\frac{1}{2\sqrt{\pi}x^{1/4}}\exp\left(-\frac{2}{3}x^{3/2}\right)\cdot\left(1-\frac{5}{48}\frac{1}{x^{3/2}}+O(x^{-3})\right),&x\to+\infty,\\[15pt]
\displaystyle\frac{1}{\sqrt{\pi}|x|^{1/4}}\left(\sin\big(\frac{2}{3}|x|^{3/2}+\frac{\pi}{4}\big)-\cos\big(\frac{2}{3}|x|^{3/2}+\frac{\pi}{4}\big)\cdot\frac{5}{36}\frac{1}{x^{3/2}}+O(x^{-3})\right),&x\to-\infty\ .
\end{cases}
\end{equation}
to obtain (with $R(\beta)<0$ for sufficiently small $\beta$)
\begin{align}
d^{j}_{nm}(\beta)&=\frac{C(\mu,\nu)}{\sqrt{4\pi J}}\Big(\frac{-1}{R(\beta)}\Big)^{1/4}e^{JS(\beta)}\cdot\Big(1+O\big(\frac{1}{\ln{\beta}}\big)\Big)\\
&=\frac{C(\mu,\nu)}{\sqrt{4\pi(n-m)}}\cdot e^{JS(\beta)}\cdot\Big(1+O\big(\frac{1}{\ln{\beta}}\big)\Big)\ .
\end{align}
We take the case of where $m<n$ as an example, in which we obtain
\begin{align}
C(\mu,\nu)&=\lim_{\beta\to0}{\sqrt{\frac{(j+n)!(j-m)!}{(j+m)!(j-n)!}}\frac{(-1)^{n-m}}{(n-m)!}\cdot\sqrt{4\pi(n-m)}\cdot\big(\frac{\beta}{2}\big)^{n-m}\cdot\exp\big(\!-\!JS(\beta)\big)}\\
&=\sqrt{\frac{(j+n)!(j-m)!}{(j+m)!(j-n)!}}\frac{(-1)^{n-m}}{(n-m)!}\cdot\sqrt{4\pi(n-m)}\cdot(\mu-\nu)^{n-m}\frac{(1+\mu)^{\textstyle\frac{J+m}{2}}(1-\nu)^{\textstyle\frac{J-n}{2}}}{(1-\mu)^{\textstyle\frac{J-m}{2}}(1+\nu)^{\textstyle\frac{J+n}{2}}}\ ,
\end{align}
which can be simplified with Stirling's formula to
\begin{equation}
C(\mu,\nu)=(-1)^{n-m}\sqrt{2}\cdot\sqrt{\frac{F(j+n)F(j-m)}{F(j+m)F(j-n)}}\cdot\frac{G(j+n)G(j-m)}{G(j+m)G(j-n)}\ ,
\end{equation}
where
\begin{equation}
F(n)=\frac{n!}{\sqrt{2\pi n}}\Big(\frac{e}{n}\Big)^{n}=1+\frac{1}{12}n+O(n^{-2}),\quad G(n)=\sqrt{\Big(1+\frac{1}{2n}\Big)^{n}}=1+\frac{3}{16n}+O(n^{-2})\ .
\end{equation}

Combined with the symmetry \cref{eqn:symmetry} of Wigner $d$-matrix, for sufficiently large $j$, we finally obtain
\begin{equation}\label{eqn:WKB-solution}
d^{j}_{nm}(\beta)=(-1)^{\max\{0,\,n-m\}}\sqrt{\frac{2}{J}}\cdot\Ai(\zeta)\Big(\frac{-\zeta}{R(\beta)}\Big)^{1/4}.
\end{equation}

\section{Asymptotic Analysis on Entropy of Wigner {\itshape \MakeLowercase{d}\,}-matrix}

In this section, we turn to calculate the entropy of Wigner $d$-matrix, which is defined as
\begin{align}\label{eqn:entropy}
\displaystyle H_{j}(\beta)&:=-\frac{1}{2j+1}\sum_{m,n=-j}^{j}\!\!{|d^{j}_{nm}(\beta)|^{2}\ln{|d^{j}_{nm}(\beta)|^{2}}}\ .
\end{align}

With the asympotic formulae \cref{eqn:Airy-asymptotic}, we have
\begin{equation}\label{eqn:norm-asymptotic}
|d_{nm}^{j}(\beta)|=\begin{cases}
\displaystyle\frac{2\sin^{2}{(JS+\pi/4)}}{\pi J\sqrt{R(\mu,\nu,\beta)}}\Big(1+O(J^{-1})\Big),&R(\beta)>0\ ,\\[10pt]
\displaystyle\frac{\exp{(2JS)}}{2\pi J\sqrt{-R(\mu,\nu,\beta)}}\Big(1+O(J^{-1})\Big),&R(\beta)<0\ .
\end{cases}
\end{equation}
It's worth noting that the boundary $R(\mu,\nu,\beta)=0$ forms an ellipse in $\mu$--$\nu$ plane. After affine transformation
\begin{equation}
\begin{cases}
\displaystyle x=(\mu+\nu)\sin{\frac{\beta}{2}}\ ,\\[10pt]
\displaystyle y=(\mu-\nu)\cos{\frac{\beta}{2}}\ ,
\end{cases}
\end{equation}
and conversion to polar coordinates, we obtain
\begin{equation}
R(r,\beta)=\sin^{2}{\beta}-r^{2}\ .
\end{equation}

According to \cref{eqn:norm-asymptotic}, we consider dividing the sum \cref{eqn:entropy} into three parts: the classically allowed region ($S_{\text{allowed}}$), the classically forbidden region ($S_{\text{forbidden}}$) and a boundary layer ($S_{\text{boundary}}$) of width $\varepsilon$.
\begin{align}
H_{j}(\beta)&=S_{\text{allowed}}+S_{\text{boundary}}+S_{\text{forbidden}}\\
&=-\frac{1}{2J}\sum_{r\le\sin{\beta}-\varepsilon}\!\!\!{|d^{j}_{nm}|^{2}\ln{|d^{j}_{nm}|^{2}}}-\frac{1}{2J}\sum_{|r-\sin{\beta}|<\varepsilon}\!\!\!{|d^{j}_{nm}|^{2}\ln{|d^{j}_{nm}|^{2}}}-\frac{1}{2J}\sum_{r\ge\sin{\beta}+\varepsilon}\!\!\!{|d^{j}_{nm}|^{2}\ln{|d^{j}_{nm}|^{2}}}\ .
\end{align}

We are to substitute the two asymptotic expressions from \label{norm-asymptotic} into $S_{\text{forbidden}}$ and $S_{\text{allowed}}$ respectively, which requires the parameter $\varepsilon$ to satisfy $\varepsilon\cdot|\partial\zeta/\partial r|\sim O(1)$. From \cref{eqn:action-func} we obtain ($\varepsilon>0$)
\begin{equation}
S(r,\beta)=-\frac{4\sqrt{2}}{3(\cos{\beta}-\cos{2\theta})\sqrt{\sin{\beta}}}\cdot\varepsilon^{3/2}+O(\varepsilon^{5/2})\ ,
\end{equation}
from which $\varepsilon\sim O(J^{-2/3})$. Thus $\varepsilon(J)$ can be set to be an infinitesimal as $J\to+\infty$, which ensures that
\begin{equation}\label{eqn:boundary-condition}
\lim_{J\to+\infty}{\frac{S_{\text{boundary}}}{S_{\text{allowed}}}}=0\ .
\end{equation}
Here, we implicitly assume that the approximate solution given by \cref{eqn:WKB-solution} remains valid in the boundary region. Although it will be shown subsequently that this assumption does not hold in general, this limitation does not compromise the validity of \cref{eqn:boundary-condition} itself.

Furthermore, given that the Airy function exhibits exponential decay in the classically forbidden region (as demonstrated in \cref{eqn:Airy-asymptotic}), it follows that
\begin{equation}
\lim_{J\to+\infty}{\frac{S_{\text{forbidden}}}{S_{\text{allowed}}}}=0\ .
\end{equation}

Therefore, the summation can be simplified into
\begin{align}
H_{j}(\beta)\approx-\frac{1}{2J}\sum_{r<\sin{\beta}}\!{|d^{j}_{nm}(\beta)|^{2}\ln{|d^{j}_{nm}(\beta)|^{2}}}=-\frac{1}{2J}\sum_{r<\sin{\beta}}\!{\frac{2\sin^{2}{(JS+\pi/4)}}{\pi J\sqrt{R(r,\beta)}}\ln{\frac{2\sin^{2}{(JS+\pi/4)}}{\pi J\sqrt{R(r,\beta)}}}}\ .
\end{align}

\subsection{Calculation of Leading and Sub-Leading Order Terms}

Considering that the function $\sin^{2}(\phi)\ln{\sin^{2}{(\phi)}}$ is $\pi$-periodic, which can be expressed as the Fourier series expansion
\begin{equation}
\sin^{2}(\phi)\ln{\sin^{2}{(\phi)}}=\frac{1-2\ln{2}}{2}+\big(\!\ln{2}-\frac{3}{4}\big)\cos{2\phi}+\sum_{k=1}^{\infty}{\left(\frac{\cos{4k\phi}}{8k^{3}-2k}+\frac{\cos{(4k+2)\phi}}{8k^{3}+12k^{2}+4k}\right)}\ .
\end{equation}
After substitution, we obtain
\begin{align}
H_{j}(\beta)\approx\frac{1}{2J}\sum_{r<\sin{\beta}}{|d^{j}_{nm}(\beta)|^{2}\big(\!\ln{(2\pi J)}-\frac{3}{2}\big)}&+\sum_{r<\sin{\beta}}{\frac{1+\ln{\sqrt{R(r,\beta)}}}{2\pi J^{2}\sqrt{R(r,\beta)}}}+\sum_{r<\sin{\beta}}{\frac{\ln{\sqrt{R(r,\beta)}}}{2\pi J^{2}\sqrt{R(r,\beta)}}\cdot\sin{(2JS)}}\notag\\
&-\sum_{r<\sin{\beta}}{\sum_{k=1}^{\infty}{\frac{2\cdot(-1)^{k}}{2\pi J^{2}\sqrt{R(r,\beta)}}\Big(\frac{\cos{(4kJS)}}{8k^{3}-2k}-\frac{\sin{\big((4k+2)JS\big)}}{8k^{3}+12k^{2}+4k}\Big)}}\ .\label{eqn:entropy-terms}
\end{align}

Now we start to calculate each term in \cref{eqn:entropy-terms}, where the first one can be obtained using the normalization condition of the Wigner $d$-matrix, namely
\begin{equation}
\frac{1}{2J}\sum_{r<\sin{\beta}}{|d^{j}_{nm}(\beta)|^{2}\big(\!\ln{(2\pi J)}-\frac{3}{2}\big)}=\ln{(2\pi J)}-\frac{3}{2}\ .
\end{equation}
And the second summation term can be transformed into an integral over continuous variables, namely
\begin{align}
\!\!\!\sum_{r<\sin{\beta}}{\!\!\frac{1+\ln{\sqrt{R(r,\beta)}}}{2\pi J^{2}\sqrt{R(r,\beta)}}}=\!\!\!\mathop{\iint}\limits_{r<\sin{\beta}}{\!\!\frac{1+\ln{\sqrt{R(r,\beta)}}}{2\pi J^{2}\sqrt{R(r,\beta)}}\ \mathrm{d}m\,\mathrm{d}n}=\!\int_{0}^{\sin{\beta}}{\frac{1+\ln{\sqrt{R(r,\beta)}}}{2\pi J^{2}\sqrt{R(r,\beta)}}\cdot\frac{2\pi r\cdot J^{2}}{\sin{\beta}}\,\mathrm{d}r}=\ln{(\sin{\beta})}-1\ .
\end{align}

The most challenging part of this section is managing the summation involving harmonic function factors. Taking the $\sin(2JS)$ term as an example, for the grid point $(\mu,\nu)$ adjacent to the grid point $(\mu_{0},\nu_{0})$, we obtain
\begin{equation}
\sin{\big(2JS(\mu,\nu)\big)}=\sin{\Big(2JS(\mu_{0},\nu_{0})+\frac{\partial S}{\partial\mu}\Big|_{(\mu_{0},\nu_{0})}\cdot 2J(\mu-\mu_{0})+\frac{\partial S}{\partial\nu}\Big|_{(\mu_{0},\nu_{0})}\cdot 2J(\nu-\nu_{0})+o(1)\Big)}\ ,
\end{equation}
where $2J(\mu-\mu_{0}),2J(\nu-\nu_{0})=\pm2$, thus the summation involving harmonic functions cannot be converted into an integral over continous variables as $J\to+\infty$.

Despite lack of means for accurate calculation, we can estimate the magnitude of the harmonic term by considering a coarse-grained grid dividing the $\mu-\nu$ plane into a number of squares, each of which contains $(2J^{\delta}+1)^{2}$ grid points, with its center denoted as $(\mu_{0},\nu_{0})$. For each square contained in the classical allowed region, we obtain
\begin{align}
&\sum_{r,s=-J^{\delta}}^{J^{\delta}}{\frac{\ln{\sqrt{R(\mu_{0}+r/J,\nu_{0}+s/J)}}}{2\pi J^{2}\sqrt{R(\mu_{0}+r/J,\nu_{0}+s/J)}}\cdot\sin{\big(2JS(\mu_{0}+r/J,\nu_{0}+s/J)\big)}}\notag\\
=\ &\frac{\ln{\sqrt{R(\mu_{0},\nu_{0})}}}{2\pi J^{2}\sqrt{R(\mu_{0},\nu_{0})}}\cdot\sum_{r,s=-J^{\delta}}^{J^{\delta}}{\left(\sin{\Big(2JS(\mu_{0},\nu_{0})+2r\cdot\frac{\partial S}{\partial\mu}\Big|_{(\mu_{0},\nu_{0})}+2s\cdot\frac{\partial S}{\partial\nu}\Big|_{(\mu_{0},\nu_{0})}\Big)+O(J^{\delta-1})}\right)}+O(J^{\delta-2})\\
=\ &\frac{\ln{\sqrt{R(\mu_{0},\nu_{0})}}}{2\pi J^{2}\sqrt{R(\mu_{0},\nu_{0})}}\cdot\frac{\displaystyle\sin{(2JS)}\sin{\big((2J^{\delta}+1)(\partial S/\partial\mu)\big)}\sin{\big((2J^{\delta}+1)(\partial S/\partial\nu)\big)}}{\sin{(\partial S/\partial\mu)}\sin{(\partial S/\partial\nu)}}\Big|_{(\mu_{0},\nu_{0})}+O(J^{3\delta-2})\ .
\end{align}
Given that $0\le\partial S/\partial\mu,\partial S/\partial\nu\le\pi$, the leading term in the summation over all squares can be bounded as
\begin{align}
&\!\sum_{\substack{(\mu_{0},\nu_{0})\\r<\sin{\beta}}}{\left|\frac{\ln{\sqrt{R(\mu_{0},\nu_{0})}}}{2\pi J^{2}\sqrt{R(\mu_{0},\nu_{0})}}\right|\cdot\frac{\displaystyle\sin{(2JS)}\sin{\big((2J^{\delta}+1)(\partial S/\partial\mu)\big)}\sin{\big((2J^{\delta}+1)(\partial S/\partial\nu)\big)}}{\sin{(\partial S/\partial\mu)}\sin{(\partial S/\partial\nu)}}\Big|_{(\mu_{0},\nu_{0})}}\notag\\
<\ &\!\sum_{\substack{(\mu_{0},\nu_{0})\\r<\sin{\beta}}}{\Big|\frac{\ln{\sqrt{R(\mu_{0},\nu_{0})}}}{2\pi J^{2}\sqrt{R(\mu_{0},\nu_{0})}}\Big|\cdot\frac{1}{\sin{(\partial S/\partial\mu)}\sin{(\partial S/\partial\nu)}}\Big|_{(\mu_{0},\nu_{0})}}\xrightarrow{J\to\infty}-\int{\frac{\ln{\sqrt{R(\mu,\nu)}}}{4\pi\sqrt{R(\mu,\nu)}}\frac{J^{1-2\delta}\,\mathrm{d}\mu\,\mathrm{d}\nu}{\sin{(\partial S/\partial\mu)}\sin{(\partial S/\partial\nu)}}}\ ,
\end{align}
which is of order $O(J^{1-2\delta})$, while the remainder is $O(J^{\delta})$ after summation. Accordingly, setting $\delta = 1/3$ yields an optimal estimate, so that the $\sin(2JS)$ term is bounded by $O(J^{1/3})$. Similar analyses can be applied to other harmonic terms.

\subsection{Analysis of Approximate Solutions in Boundary Regions}

At the end of this section, we verify the validity of the approximate solution \cref{eqn:WKB-solution} in the boundary layer. In this case $R(\beta)\to0$, thereby the discarded term in \cref{eqn:Airy-eqn} cannot be ignored, as it becomes
\begin{equation}
\frac{1}{J^{2}}\cdot\frac{\delta(\beta)}{4R^{3}(\beta)}\approx\frac{1}{J^{2}x^{3}}\cdot\frac{\delta(\beta)}{32\sin^{3}{\beta}}\to\infty
\end{equation}
with $x=\sin{\beta}-r\sim o(J^{-2/3})$. We consider the Taylor series near $\zeta=0$ and retain the first term, thus obtain
\begin{equation}\label{eqn:Taylor-Airy}
\frac{\mathrm{d}^{2}h}{\mathrm{d}\zeta^{2}}=\left(1+\frac{1}{J^{2}}\cdot\frac{\delta(\beta)}{4R^{3}(\beta)}\right)\zeta\cdot h(\zeta)=\left(1+\frac{3(5+\cos{2\beta_{\pm}}-6\cos{\beta_{\pm}}\cos{2\phi})}{64J^{2/3}\zeta^{2}\big((\cos{\beta_{\pm}}-\cos{2\phi})^{2}\sin{\beta_{\pm}}\big)^{2/3}}+O\big(\frac{1}{J^{4/3}\zeta}\big)\right)\zeta\cdot h(\zeta)\ ,
\end{equation}
where $\beta_{\pm}$ is the turning point in consideration, and
\begin{equation}
\phi=\arctan{\Big(\frac{1}{\tan{(\beta_{\pm}/2)}}\frac{\mu-\nu}{\mu+\nu}\Big)}\ .
\end{equation}
From \cref{eqn:Taylor-Airy}, one could find the uniform approximation method invalidates only if $\zeta\sim O(J^{-1/3})$, or equivalently $\varepsilon\sim O(J^{-1})$. Hence, it is reasonable to take the approximate solution \cref{eqn:airy} when calculating $S_{\text{boundary}}$ in the boundary layer.

Combining the previous analysis, we obtain the asymptotic expression for the entropy of the Wigner $d$-matrix in the macroscopic limit $j \to \infty$: 
\begin{equation}\label{eqn:semi-approx-supp}
H_{j}(\beta)=\ln{(2\pi J)}+\ln{|\sin{\beta}|}-\frac{5}{2}+o(1)\ .
\end{equation}

\section{Analysis near the singularity $\beta\to n\pi\,(n\in\mathbb{Z})$}

According to the previous discussion, the asymptotic expression in \cref{eqn:semi-approx-supp} diverges as $\beta\to n\pi$ for $n\in\mathbb{Z}$. Such divergence arises because the asymptotic analysis is valid only in the regime where $S_{\text{allowed}}\gg S_{\text{boundary}}$, or equivalently, $|\beta-n\pi|\gg J^{-2/3}$. As $\beta$ approaches the singularities, the classically allowed region contracts while the boundary region becomes increasingly dominant. Consequently, the WKB approximation and the resulting asymptotic expression \cref{eqn:semi-approx-supp} breaks down.

Owing to the inherent symmetry of $d_{nm}^{j}(\beta)$, it is sufficient to consider the limit $\beta \to 0$, without loss of generality. In such regime, an alternative analytical approach is to employ Wigner's series expansion for $d_{nm}^{j}(\beta)$, as presented in Ref.\,\cite[Chap.\,15]{Wig59a}, and retaining only the leading-order term, from which we obtain
\begin{equation}
d^{j}_{nm}(\beta)=\sqrt{\frac{(j+a)!(j-b)!}{(j+b)!(j-a)!}}\frac{(-1)^{a-b}}{(a-b)!}\cdot\big(\frac{\beta}{2}\big)^{a-b}+O(\beta^{a-b+2})\ .
\end{equation}
where $a=\max(m,n),b=\min(m,n)$.

Consider the summation of $|d_{nm}^{j}(\beta)|^{2} \ln |d_{nm}^{j}(\beta)|^{2}$, grouped according to the index difference $k=|m-n|$. For $k=0$, we obtain
\begin{align}
S_{0}&:=-\frac{1}{2j+1}\sum_{n=-j}^{j}{|d_{nn}^{j}(\beta)|^{2}\ln|d_{nn}^{j}(\beta)|^{2}}\\
&=-\frac{1}{2j+1}\sum_{n=-j}^{j}{\Big(1-\frac{\beta^{2}}{2}\big(j(j+1)-n^{2}\big)\Big)\ln\Big(1-\frac{\beta^{2}}{2}\big(j(j+1)-n^{2}\big)\Big)}\\
&=\frac{\beta^{2}}{2(2j+1)}\sum_{n=-j}^{j}{\big(j(j+1)-n^{2}\big)}+O(\beta^{2})=\frac{\beta^{2}}{3}\big(J^{2}-\frac{1}{4}\big)+O(\beta^{2})\ .
\end{align}
For $k\ne0$, we obtain
\begin{align}\label{eqn:case-k}
S_{k}&:=-\frac{1}{2j+1}\sum_{n=-j}^{j-k}{|d_{n,n+k}^{j}(\beta)|^{2}\ln|d_{n,n+k}^{j}(\beta)|^{2}}-\frac{1}{2j+1}\sum_{n=-j+k}^{j}{|d_{n,n-k}^{j}(\beta)|^{2}\ln|d_{n,n-k}^{j}(\beta)|^{2}}\notag\\
&=-\sum_{n=-j}^{j-k}{\frac{2}{2j+1}\big(\frac{\beta^{2k}}{2^{2k}(k!)^{2}}\frac{(j+n+k)!}{(j+n)!}\frac{(j-n)!}{(j-n-k)!}\big)\ln{\big(\frac{\beta^{2k}}{2^{2k}(k!)^{2}}\frac{(j+n+k)!}{(j+n)!}\frac{(j-n)!}{(j-n-k)!}\big)}}+O(\beta^{2})\ .
\end{align}
Applying the mean value inequality, and defining $\beta = J^{-1-\delta}\,(\delta>0)$ for simplicity, we obtain
\begin{equation}
\frac{\beta^{2k}}{2^{2k}(k!)^{2}}\frac{(j+n+k)!}{(j+n)!}\frac{(j-n)!}{(j-n-k)!}\le\frac{(j+1/2)^{2k}\beta^{2k}}{2^{2k}(k!)^{2}}=\frac{J^{-2k\delta}}{2^{2k}(k!)^{2}}\le\frac{1}{4}\ ,
\end{equation}
from which an upper bound for $S_{k}$ follows: 
\begin{align}
S_{k}&\le-\sum_{n=-j}^{j-k}{\frac{2}{2j+1}\frac{J^{-2k\delta}}{2^{2k}(k!)^{2}}\ln{\frac{J^{-2k\delta}}{2^{2k}(k!)^{2}}}}+O(\beta^{2})=-\frac{(2J-k)}{J}\frac{J^{-2k\delta}}{2^{2k}(k!)^{2}}\ln{\frac{J^{-2k\delta}}{2^{2k}(k!)^{2}}}+O(\beta^{2})\ .
\end{align}

Therefore, 
\begin{align}
H_{j}(\beta)&=\sum_{k=0}^{2j}{S_{k}}\le\frac{\beta^{2}}{3}\big(J^{2}-\frac{1}{4}\big)-\sum_{k=1}^{2J-1}{\frac{(2J-k)}{J}\frac{J^{-2k\delta}}{2^{2k}(k!)^{2}}\ln{\frac{J^{-2k\delta}}{2^{2k}(k!)^{2}}}}+O(\beta^{2})\\
&\le\frac{\beta^{2}}{3}\big(J^{2}-\frac{1}{4}\big)-\sum_{k=1}^{\infty}{2\cdot\frac{J^{-2k\delta}}{2^{2k}(k!)^{2}}\ln{\frac{J^{-2k\delta}}{2^{2k}(k!)^{2}}}}+O(\beta^{2})\\
&\le\frac{\beta^{2}}{3}\big(J^{2}-\frac{1}{4}\big)-\sum_{k=1}^{\infty}{2\big(\frac{J^{-k\delta}}{2^{k+1}k!}+\frac{J^{-2k\delta}}{2^{2k}(k!)^{2}}\big)}+O(\beta^{2})=\frac{J^{-\delta}}{4}+O(J^{-2\delta})\ ,
\end{align}
from which it follows that the leading order of $H_{j}(\beta)$ does not exceed $O(1)$.

\section{Ideal negative measurements for higher-order temporal correlations}

Ideal negative measurement (INM) protocols discussed in the main text can be naturally extended to measure multi-time joint probabilities.  In this section, we focus on the case of three-time joint probabilities for illustration. Each subsection first briefly reviews the implementation for the two-time senario, and then outlines a direct generalization to the three-time scenario. These examples are given for clarity; the same principles apply to correlations of arbitrary order.

\subsection{Ancilla-assisted schemes}

The invasiveness of quantum measurements arises from the fact that projective measurements inevitably collapse the system state onto an eigenstate associated with the measurement outcome. Thus, unless the density matrix happens to be diagonal in the measurement basis at the time of measurement, the act of measurement will necessarily disturb the system state. A viable strategy to mitigate such disturbance is to avoid projective readout at intermediate times and instead couple the system to an ancilla that becomes entangled with the system in the measurement basis.

Following the approach proposed by Katiyar {\itshape et al}. \cite{KSG+12}, consider measuring a pure stat
\begin{equation}
\ket{\psi_{\text{S}}} = c_{0}\ket{0_{\text{S}}} + c_{1}\ket{1_{\text{S}}} + c_{2}\ket{2_{\text{S}}},
\end{equation}
with respect to the projector $P_{0} = \ket{0}\bra{0}$. We introduce an ancillary qubit initially prepared in $\ket{\psi_{A}} = \ket{0_{A}}$. At time $t_{i}$, we apply to the composite system the controlled gate
\begin{equation}
\mathrm{CG}_{0}=\ket{0_{\text{S}}}\bra{0_{\text{S}}}\otimes I_{A}+\ket{1_{\text{S}}}\bra{1_{\text{S}}}\otimes X_{A}+\ket{2_{\text{S}}}\bra{2_{\text{S}}}\otimes X_{A},
\end{equation}
which transforms the joint state into
\begin{equation}
\ket{\psi_{SA}}=c_{0}\ket{0_{\text{S}},0_{A}}+c_{1}\ket{1_{\text{S}},1_{A}}+c_{2}\ket{2_{\text{S}},1_{A}}.
\end{equation}
Hence the ancilla records information about the system’s state at time $t_{i}$ without projecting the system itself, permitting all physical measurements to be deferred to the final time. A joint measurement on the system andancilla then yields the correct joint probabilities.

This construction generalizes straightforwardly to multi-time joint probabilities. For example, to measure the probability $p(q_{1}=0, q_{2}=0, q_{3}=0)$, one applies at times $t_{1}$ and $t_{2}$ the controlled gates
\begin{align}
\mathrm{CG}_{0}^{(1)}&=\ket{0_{\text{S}}}\bra{0_{\text{S}}}\otimes I_{\text{A1}}\otimes I_{\text{A2}}+\ket{1_{\text{S}}}\bra{1_{\text{S}}}\otimes X_{\text{A1}}\otimes I_{\text{A2}}+\ket{2_{\text{S}}}\bra{2_{\text{S}}}\otimes X_{\text{A1}}\otimes I_{\text{A2}},\\
\mathrm{CG}_{0}^{(2)}&=\ket{0_{\text{S}}}\bra{0_{\text{S}}}\otimes I_{\text{A1}}\otimes I_{\text{A2}}+\ket{1_{\text{S}}}\bra{1_{\text{S}}}\otimes I_{\text{A1}}\otimes X_{\text{A2}}+\ket{2_{\text{S}}}\bra{2_{\text{S}}}\otimes I_{\text{A1}}\otimes X_{\text{A2}},
\end{align}
and then performs at time $t_{3}$ the projective measurement
\begin{equation}
\Pi_{000}=\ket{0_{\text{S}}}\bra{0_{\text{S}}}\otimes\ket{0_{\text{A1}}}\bra{0_{\text{A1}}}\otimes\ket{0_{\text{A2}}}\bra{0_{\text{A2}}}.
\end{equation}
This procedure yields the correct three-time probability distribution.

The scheme above addresses dichotomic measurements. For multi-outcome sharp measurements one can introduce controlled gates of the form (omitting the identity action on irrelevant subspaces for simplicity)
\begin{equation}\label{eqn:control-gate}
\mathrm{CG}^{(i)}=\ket{0_{\text{S}}}\bra{0_{\text{S}}}\otimes\ket{0_{\text{A}i}}\bra{0_{\text{A}i}}+\ket{1_{\text{S}}}\bra{1_{\text{S}}}\otimes\ket{1_{\text{A}i}}\bra{0_{\text{A}i}}+\ket{2_{\text{S}}}\bra{2_{\text{S}}}\otimes\ket{2_{\text{A}i}}\bra{0_{\text{A}i}}+\cdots.
\end{equation}
One applies these controlled gates sequentially at the measurement times $t_{i}$, and finally at time $t_{n}$ performs the projective measurement
\begin{equation}\Pi_{q^{(1)}q^{(2)}\cdots q^{(n)}}=\ket{q^{(n)}_{\text{S}}}\bra{q^{(n)}_{\text{S}}} \otimes \ket{q^{(1)}_{\text{A1}}}\bra{q^{(1)}_{\text{A1}}} \otimes\ket{q^{(2)}_{\text{A2}}}\bra{q^{(2)}_{\text{A2}}}\otimes\cdots.
\end{equation}
In principle, the observed statistics then directly yield the joint distribution $p(q^{(1)},q^{(2)},\ldots,q^{(n)})$.

\begin{figure}[ht]
\centering
\includegraphics[width=0.52\textwidth]{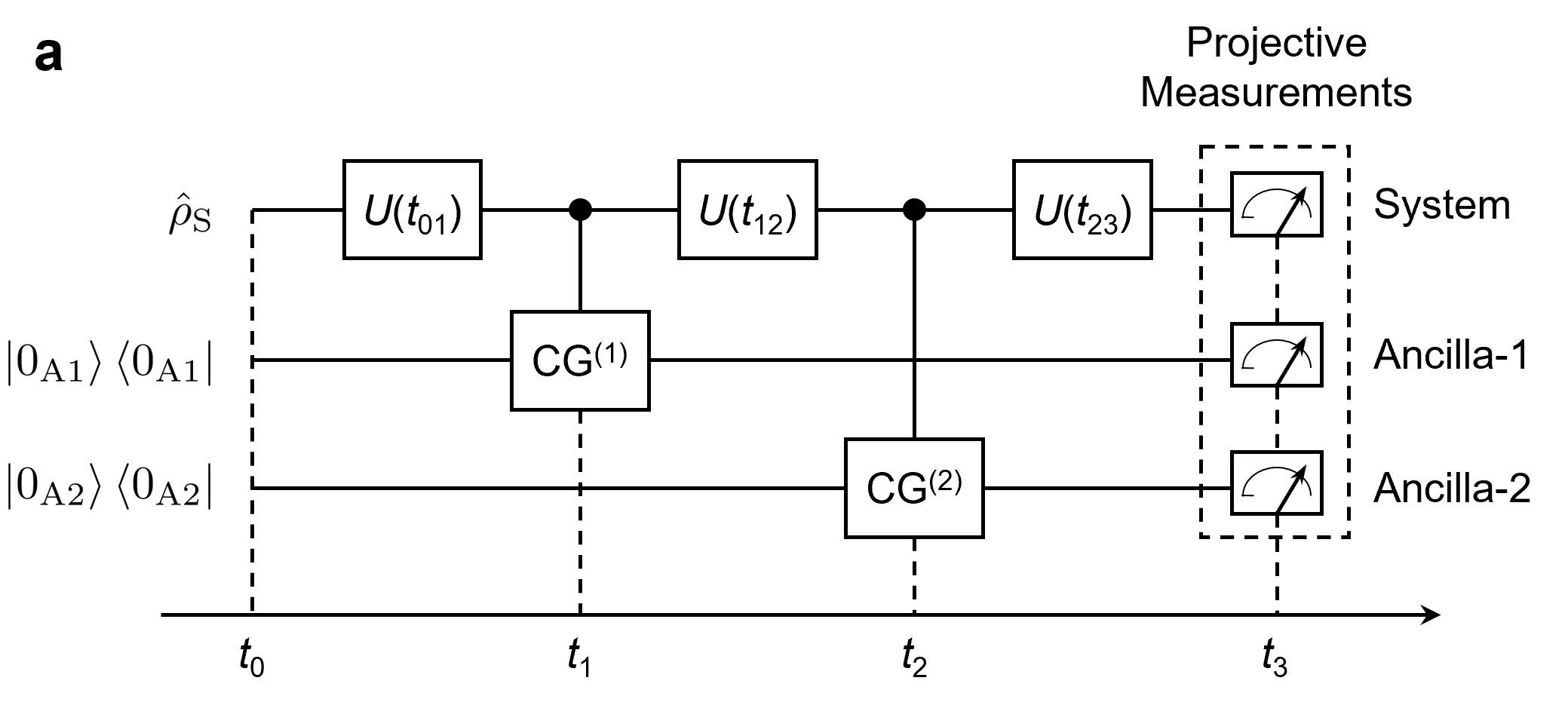}
\includegraphics[width=0.47\textwidth]{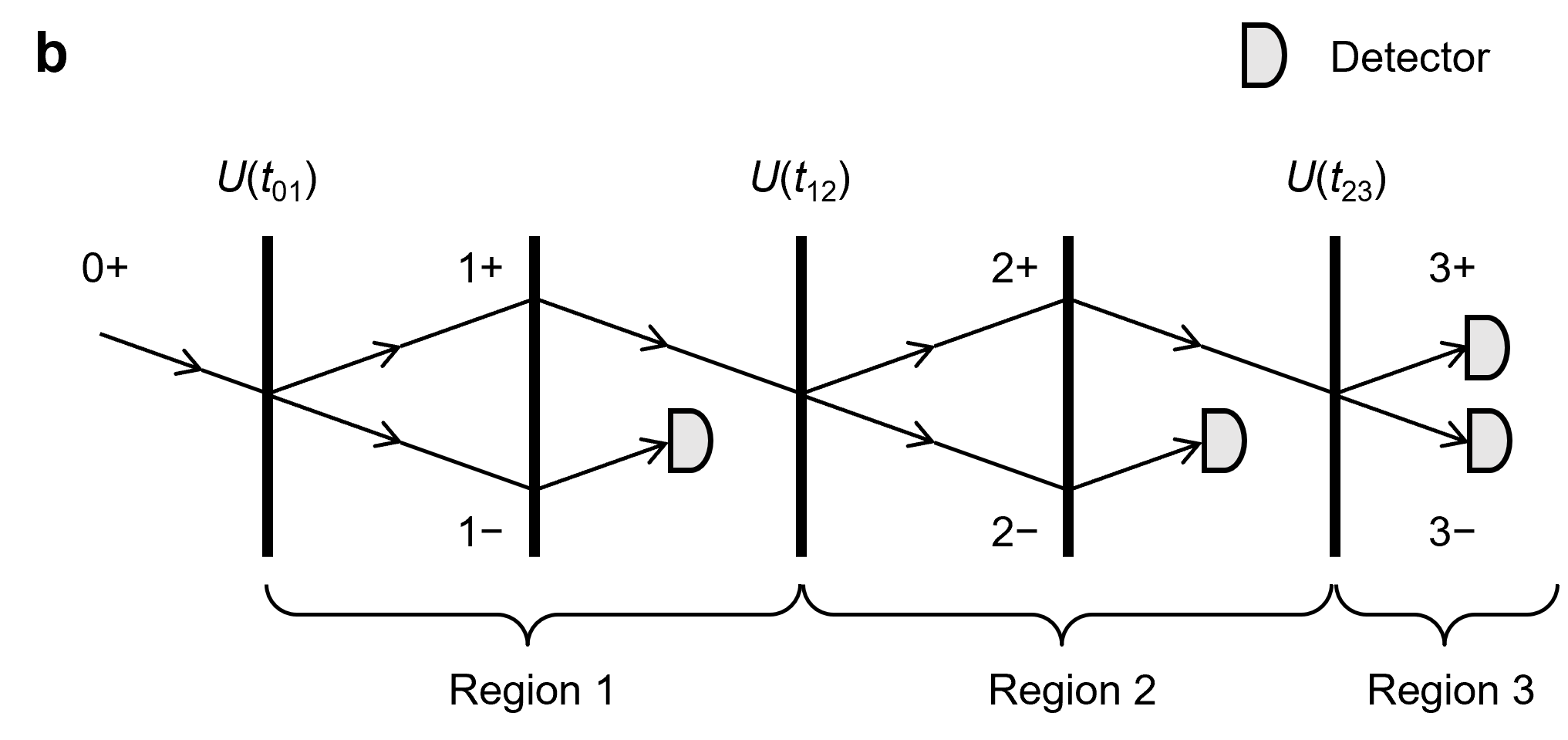}
\caption{{\bfseries Schematic representation of the ideal negative measurement protocols.} 
(a) Measurement of three-time joint probabilities via ancilla-assisted schemes. Ancilla-1 and Ancilla-2 are auxiliary systems with the same dimension as the system, both initialized in the state $\ket{0_{\text{A}i}}\bra{0_{\text{A}i}}$. $\text{CG}^{(1)}$ and $\text{CG}^{(2)}$ denote the controlled quantum gates defined in \cref{eqn:control-gate}, which implement unitary transformations on the ancilla conditioned on the system state. Projective measurements at intermediate time points are replaced by controlled gates, with a single projective readout performed at the final time $t_{3}$.  
(b) Measurement of three-time joint probabilities via interferometric null-result schemes. The depicted setup directly measures only the parameter $N_{1+2+3\pm}$; the remaining six parameters $N_{1+2-3\pm}$, $N_{1-2+3\pm}$, and $N_{1-2-3\pm}$ are obtained by switching the detector in region 1 and 2.}
\label{fig:non-invasive}
\end{figure}

\subsection{Interferometric null-result schemes}

A conceptual concern regarding ancilla-assisted schemes is that their claimed non-invasiveness relies on accepting the validity of quantum mechanics, at least insofar as the quantum description of the controlled gates is accurate. Indeed, any experimental protocol that purports to be non-invasive must invoke some underlying physical theory to justify this claim; hence, non-invasiveness can only be assumed based on auxiliary assumptions about the validity of that theory. It is therefore impossible to devise an experimental procedure that is unconditionally non-invasive. Leggett \cite{Leg08} emphasized this point and further argued that adopting locality as an additional assumption within a classical worldview is a relatively modest and acceptable assumption.

Concretely, in a Mach--Zehnder interferometer, monitoring one arm and observing a non-click event allows the experimenter to infer that the particle traveled through the other arm; under the locality assumption, this inference implies that the system's state remained undisturbed. Within a classical worldview, where measurements are assumed a priori to be local, such a null-result (non-detection) measurement is naturally considered non-invasive.

As a concrete example, we extend the scheme of Kreuzgruber {\itshape et al.} \cite{KWG+24} to the case of three measurement times, as illustrated in \cref{fig:non-invasive}b. We focus on the particle or light intensity detected in Region 3, because any particles detected there must have bypassed the detectors in Region 1 and Region 2. The corresponding paths can therefore be recorded as $(1+,2+,3\pm)$, and the counts of the two detectors are denoted as $N_{1+2+3\pm}$. By switching the detector in Region 1 between the $1+$ and $1-$ paths, and the detector in Region 2 between the $2+$ and $2-$ paths, the remaining six measurement outcomes, $N_{1-2+3\pm},N_{1+2-3\pm},N_{1-2-3\pm}$, can be obtained. The three-time joint probability distribution is thus given by
\begin{equation}
p(1x,2y,3z)=\frac{N_{1x,2y,3z}}{\displaystyle\sum_{i,j,k=\pm}^{\ }{N_{1i,2j,3k}}}\qquad(x,y,z=\pm)
\end{equation}

\end{document}